\documentclass[10pt,a4paper]{article}
\usepackage{jheppub_kim}
\usepackage{pdflscape}
\usepackage{amsmath}
\usepackage{amssymb}
\usepackage{dcolumn}
\usepackage{bm}
\usepackage{color}
\usepackage{epsfig}
\usepackage{amsfonts}
\usepackage{graphicx}
\usepackage{subfigure}
\usepackage{dcolumn}

\begin{document}
\title{Quantum Corrections to the Thermodynamics of Black Branes}
\author[a]{Behnam Pourhassan,}
\author[b]{Mir Faizal}

\affiliation[a] {School of Physics, Damghan University, Damghan, 3671641167, Iran.}
\affiliation[b] {Department of Physics and Astronomy, University of Lethbridge, Lethbridge, Alberta, T1K 3M4, Canada.}
\affiliation[b] {Irving K. Barber School of Arts and Sciences, University of British Columbia, Kelowna, British Columbia,
V1V 1V7, Canada.}
\affiliation[b] {Canadian Quantum Research Center 204-3002 32 Ave Vernon, BC V1T 2L7 Canada.}

\emailAdd{b.pourhassan@du.ac.ir}
\emailAdd{mirfaizalmir@googlemail.com}

\abstract{In this paper we study the thermodynamics of black branes at quantum scales. We analyze both perturbative and non-perturbative
corrections to the thermodynamics of such black branes. It will be observed that these corrections will modify the relation between
the entropy and area of these black branes. This will in turn modify their specific heat, and thus their stability. So, such
corrections can have important consequences for the stability of black branes at quantum scales. We also analyze the
effect of these perturbative and non-perturbative quantum corrections on various other thermodynamic quantities. Then, we
obtain the metric for the quantum corrected geometry for black branes. }

\keywords{Quantum correction; Thermodynamics; String theory}

\maketitle
\section{Introduction}
The holographic principle states that the information contained in a region of space scales with its area  \cite{4a, 5a}. This has been motivated from black hole thermodynamics, as the entropy of a black object is equal to a quarter of its area  \cite{1a, 2a}. Now the black holes are maximum entropy objects, so the maximum entropy of a region scales with its area.   It has been argued that the quantum fluctuations at the Planck scale can modify the relation between the area and entropy of a black hole, and these can again be analyzed using the holographic principle  \cite{6a, 7a}. It may be noted in such modifications, the entropy still scales with some function of the area. However, the exact function of the area would be modified by the quantum corrections to the geometry of space-time.
As AdS/CFT correspondence  relates the string theory on AdS space-time to the conformal field theory on its boundary, it is one of the most important realizations of the holographic principle. Thus,  it has been used to analyze perturbative corrections to the entropy of  AdS black holes  \cite{18, 18a, 18b, 18c, 18d}. The extremal limit of black holes has also been used to obtain such perturbative corrections \cite{19, 19a}.  It has been argued that such corrections can be obtained from the density of microstates of a conformal field theory  \cite{Ashtekar}. The quantum correction to the  Cardy formula for black holes has also been used to obtain such perturbative corrections \cite{Govindarajan}. Such corrections to the entropy of a black hole have also been calculated using the  Rademacher expansion   \cite{29}. Thus, the perturbative quantum corrections to the entropy of a black hole have been thoroughly studied using various different approaches.

It may be noted that such perturbative corrections to the entropy of black holes can also be obtained from thermal fluctuations  \cite{32, 32a, 32b, 32c, 32d}. It is possible to relate these thermal fluctuations to quantum fluctuations using the  Jacobson formalism    \cite{gr12}. This is because in the Jacobson formalism, the geometry of space-time is an emergent structure, which emerges from thermodynamics     \cite{gr12}. So, the quantum fluctuations in the geometry of space-time can be obtained from the thermal fluctuations in thermodynamics  \cite{gr14, gr16}. Thus, perturbative quantum corrections to various black holes have also been obtained from thermal fluctuations   \cite{40a, 40b, 40c, 40d}. Thus, it is possible to first obtain thermal fluctuations to the thermodynamics of various different black holes, and then use the Jacobson formalism    \cite{gr12} to obtain quantum corrections to the geometry of those black holes. It may be noted that  it has been proposed that the black hole thermodynamics can be analyzed using statistical mechanics of microstates of black holes \cite{point4, point5}. Thus, it is possible to view these the quantum corrections to the geometry occurring from the thermal fluctuations of these mircostates of geometry of black holes.

It may be noted that when a black object is very large, the quantum corrections to its metric can be neglected. Furthermore, we can also neglect the thermal fluctuations to its thermodynamics. However, as this black object reduces in size, we cannot neglect its quantum corrections. Due to this reduction in its size, its temperature also increases, and we cannot also neglect the thermal fluctuations to its thermodynamics.
At a sufficiently small size, these corrections can be analyzed as perturbative corrections to the original thermodynamics. However, when the size of the black hole is close to the  Planck scale, non-perturbative corrections become important, and cannot be neglected.   It has been proposed that such non-perturbative corrections can be expressed as the exponential function of the original entropy, for any black object   \cite{2007.15401}. Such non-perturbative exponential corrections have also been obtained using  Kloosterman sums \cite{Dabholkar}.  These corrections are obtained using    AdS/CFT correspondence for  massless supergravity fields near the horizon \cite{ds12, ds14}.  So, these non-perturbative corrections  \cite{2007.15401, Dabholkar} have  been   motivated from string theoretical effects \cite{ds12, ds14}. Thus, it is important to analyze their effects on  geometries motivated by string theory, such as black branes  \cite{ca15, ca16, da12, da14}.

It has been demonstrated that perturbative corrections to the entropy of black holes can have important consequences for their thermodynamic stability  \cite{32, 32a, 32b, 32c, 32d}. So, it is expected that non-perturbative corrections would produce interesting modifications to the thermodynamic stability of black objects.   It is also known that black branes have interesting thermodynamic behavior \cite{9412184, Peet:2000hn}. So, we will study the effect of both the perturbative and non-perturbative corrections on the thermodynamics of such black branes. As the thermodynamic stability of black branes can be modified by such quantum corrections, we will analyze the effects of such quantum corrections on the stability of such black branes. It will be observed that both perturbative and non-perturbative corrections can produce important modifications to the thermodynamic behavior of black branes.
\section{Black  Brane}
Now it is possible to analyze a   black  brane solution. Such black branes   have translational symmetry in  spatial directions.
Now we can write explicitly the such black brane solution, which will be a solution to a supergravity action.
Now for $p\leq6$,  we can write a suitable black brane metric as  \cite{9412184, Peet:2000hn},
\begin{equation}\label{metric}
ds^{2}=\frac{1}{\sqrt{D_{p}(r)}}\left(-K(r) dt^{2}+dx_{\|}^{2}\right)+\sqrt{D_{p}(r)}\left(\frac{dr^{2}}{K(r)}+r^{2}d\Omega_{8-p}^{2}\right) ,
\end{equation}
where $D_{p}(r)$ is defined as
\begin{equation}\label{metric f}
D_{p}(r)=1+\left(\frac{r_{H}}{r}\right)^{7-p}\sinh^{2}\beta,
\end{equation}
and $K(r)$ is defined as
\begin{equation}\label{metric ff}
K(r)=1-\left(\frac{r_{H}}{r}\right)^{7-p}.
\end{equation}
For the boost parameter $\beta$, we can  write $\sinh^{2}\beta$ as
\begin{equation}\label{boost}
\sinh^{2}\beta=-\frac{1}{2}+\sqrt{\frac{1}{4}+\left(c_{p}g_{s}N\left(\frac{l_{s}}{r_{H}}\right)^{7-p}\right)^{2}},
\end{equation}
where $2 \pi l_s^2$ is the  string tension, $g_s$ is the string coupling constant, and  $c_p = (2 \pi)^{5-p}\Gamma [(7-p)/2]$.
This black brane solution is the classical solution to the supergravity approximation to string theory \cite{9412184, Peet:2000hn}.   It is interesting to go beyond the supergravity approximation and analyze the quantum corrections to such a solution. Even though it is not known how such non-perturbative quantum corrections can be explicitly obtained, it is known that they would  modify the relation between the area and entropy of black branes. In fact, the perturbative \cite{40a, 40b, 40c, 40d} and non-perturbative  \cite{2007.15401, Dabholkar} corrections to the relation between area and entropy for any black object have the same functional dependence  on the original area and temperature of that black object.

Thus, we will use this functional form for such perturbative and non-perturbative corrections for such a black brane, and thus analyze the effect of such quantum corrections on the stability of such a black brane.
Now it is known that the perturbative quantum corrections correct the relation between  area entropy $S = A/4$ as  \cite{40a, 40b, 40c, 40d}
\begin{equation}
S_{c, per}=\alpha\ln{\frac{A}{4G}}+\frac{4G \lambda }{A}+\eta +\cdots,
\end{equation}
where $\alpha$ and $\lambda$ are constants.
Furthermore, it has been argued that  the non-perturbative correction to the relation between area and entropy is given by \cite{2007.15401, Dabholkar}
\begin{equation}\label{corrected entropy}
S_{c, non-per}= \eta e^{-\frac{A}{4G}}
\end{equation}
Now, the total entropy can be expressed as a sum of the original entropy with  both perturbative and non-pertubative corrections
\begin{equation}\label{corrected entropy}
S_{BH}=\frac{A}{4G}+\alpha\ln{\frac{A}{4G}}+\frac{4G \lambda}{A}+\eta e^{-\frac{A}{4G}}+\cdots
\end{equation}
It may be noted that the value of $(\alpha,    \lambda, \eta )$ control the relative strength of such corrections. To explicitly derive their values, we need to calculate explicit quantum corrections to the supergravity action, and then obtain the corrections to the geometry from such corrected supergravity action. It is not known how to obtain such non-perturbative corrections to supergravity approximation to string theory, and so it is not possible to explicitly calculate them using low energy effective field theory. However, it has been argued that both perturbative \cite{40a, 40b, 40c, 40d}    and non-perturbative \cite{2007.15401, Dabholkar}    quantum corrections to the entropy of any black object is a well defined function of the original area. In fact, as this functional form of the corrected entropy is known it can be used to analyze the quantum corrections to the stability   of black branes, without explicitly calculating the corrections to the low   energy effective action.   However, in this analysis, we are not able to fix the value of the parameter space of these corrections,   $(\alpha,    \lambda, \eta )$, and so here we will analyze the stability for general values of these parameters.
\section{Thermodynamics}
It is known that in the Jacobson formalism  \cite{gr12}  quantum fluctuations of its geometry. can be obtained from  thermal fluctuations to the thermodynamics  \cite{gr14, gr16}.  So, we will first analyze corrections to the thermodynamics of black branes.
 We can use this general functional form for quantum corrected entropy and analyze the effect of such quantum corrections on the thermodynamic stability of the black brane. Now  we  can write the   temperature  of the black brane as
\begin{equation}\label{T}
T=\frac{7-p}{4\pi r_{H}\cosh\beta}.
\end{equation}
Now we can expression the corrected entropy of this system in terms of the original entropy, and write it as
\begin{eqnarray}\label{S}
S_{BH}&=&\frac{\Omega_{8-p}}{4G_{10-p}}r_{H}^{8-p}\cosh\beta+\alpha\ln{\frac{\Omega_{8-p}}{4G_{10-p}}r_{H}^{8-p}\cosh\beta}\nonumber\\
&+&\frac{4G_{10-p}\lambda}{\Omega_{8-p}r_{H}^{8-p}\cosh\beta}+\eta e^{-\frac{\Omega_{8-p}}{4G_{10-p}}r_{H}^{8-p}\cosh\beta}.
\end{eqnarray}
It may be noted that we have not corrected the original temperature. This is because it is known that both the perturbative and non-perturbative corrections to the entropy of any black object can be expressed in terms of the original temperature and entropy of that black object. In fact, it has been even demonstrated  using the AdS/CFT correspondence that corrections to the entropy of a black hole can be expressed as functions of the original entropy and temperature of that black hole \cite{x1, x2, x21, x4}.

\begin{figure}[h!]
\begin{center}$
\begin{array}{cccc}
\includegraphics[width=50 mm]{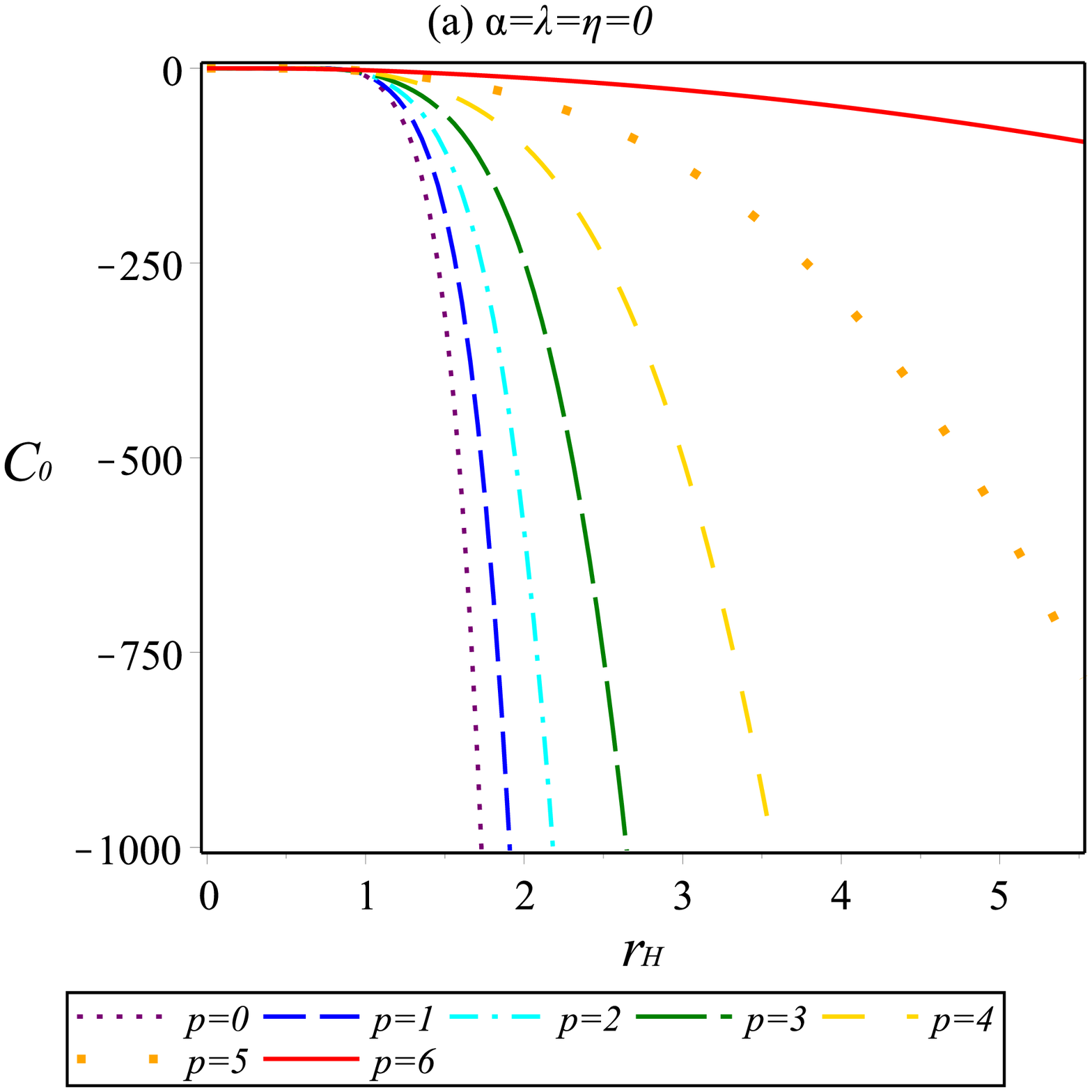}\includegraphics[width=50 mm]{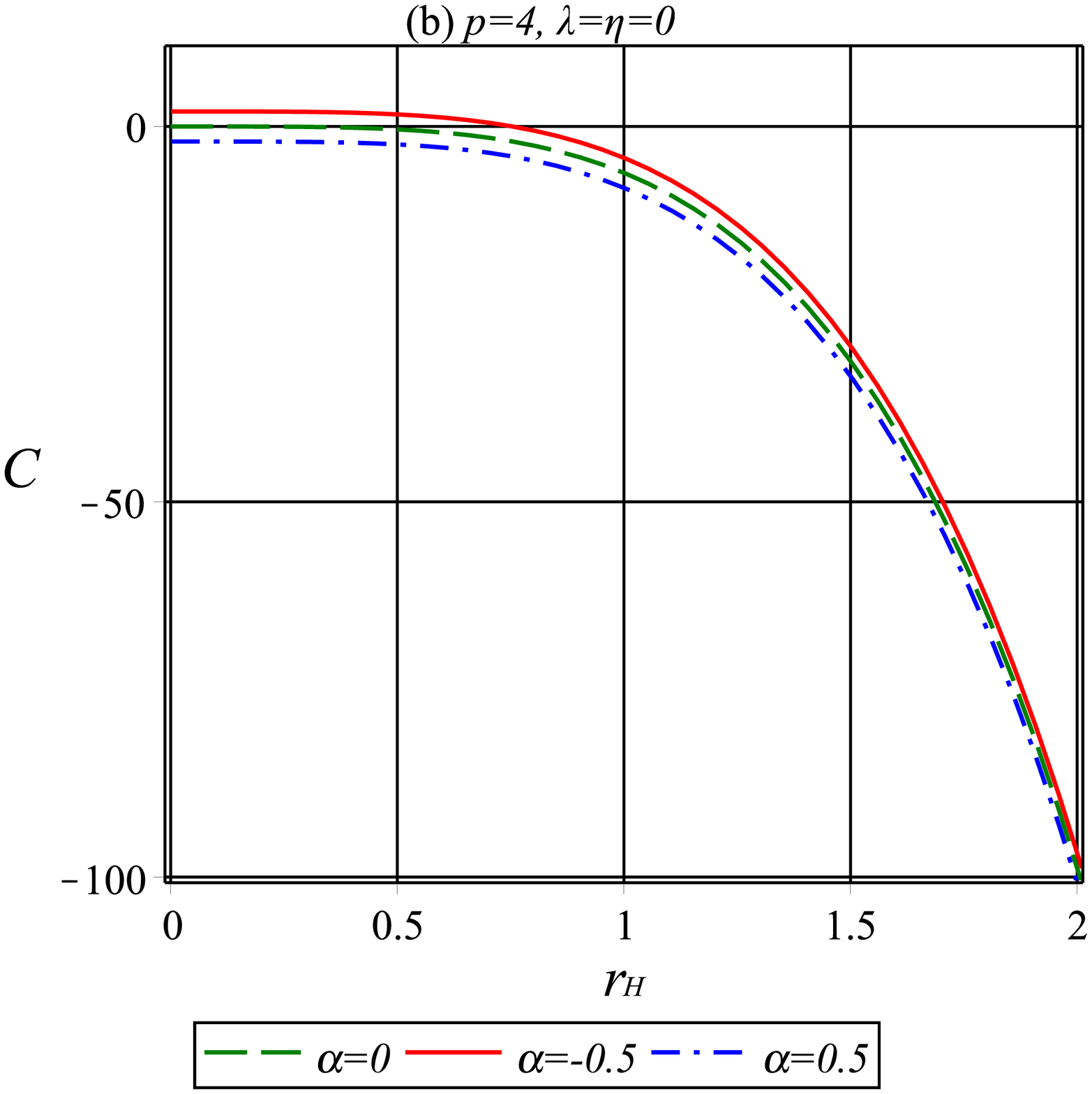}\includegraphics[width=50 mm]{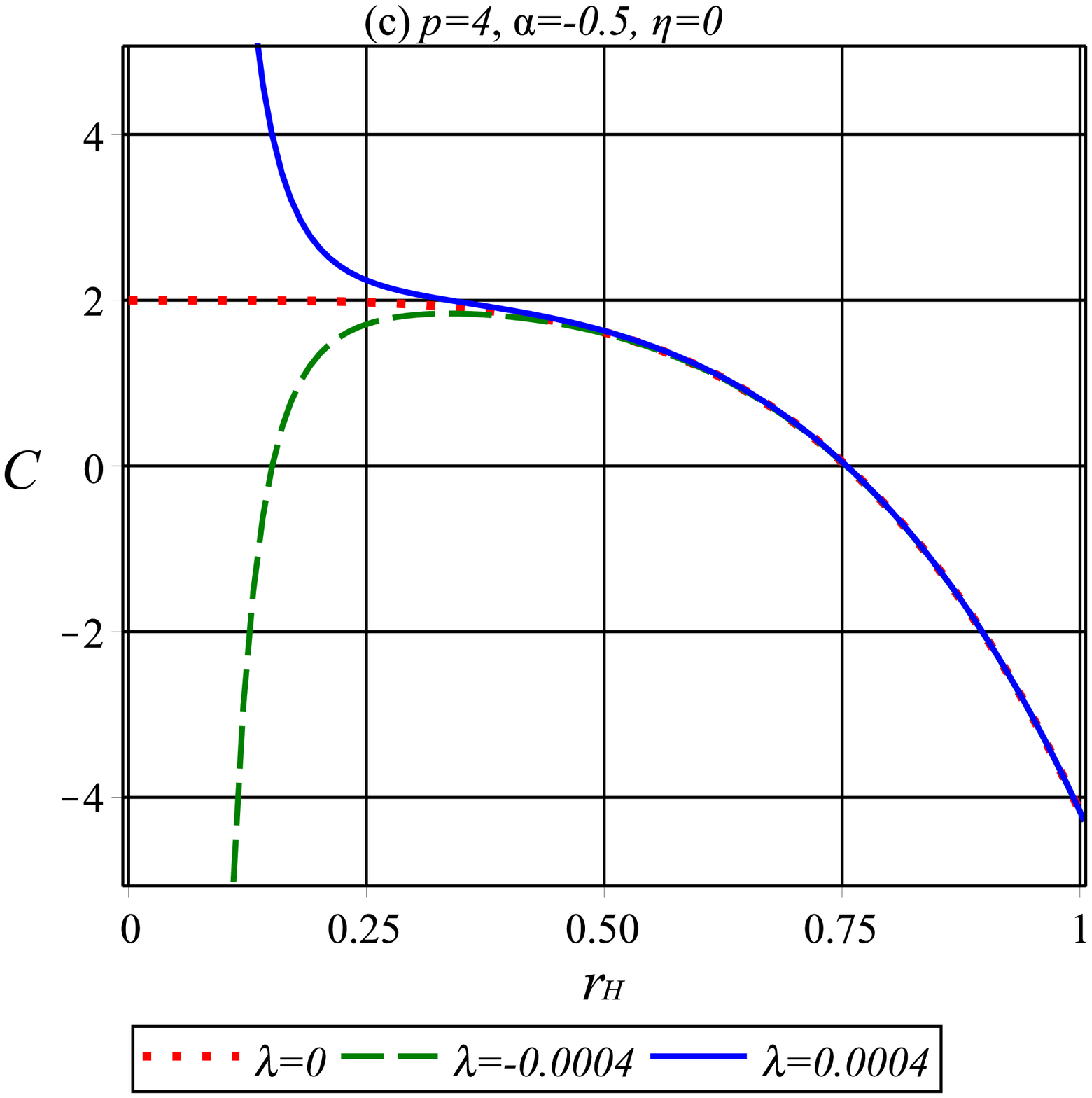}\\
\includegraphics[width=50 mm]{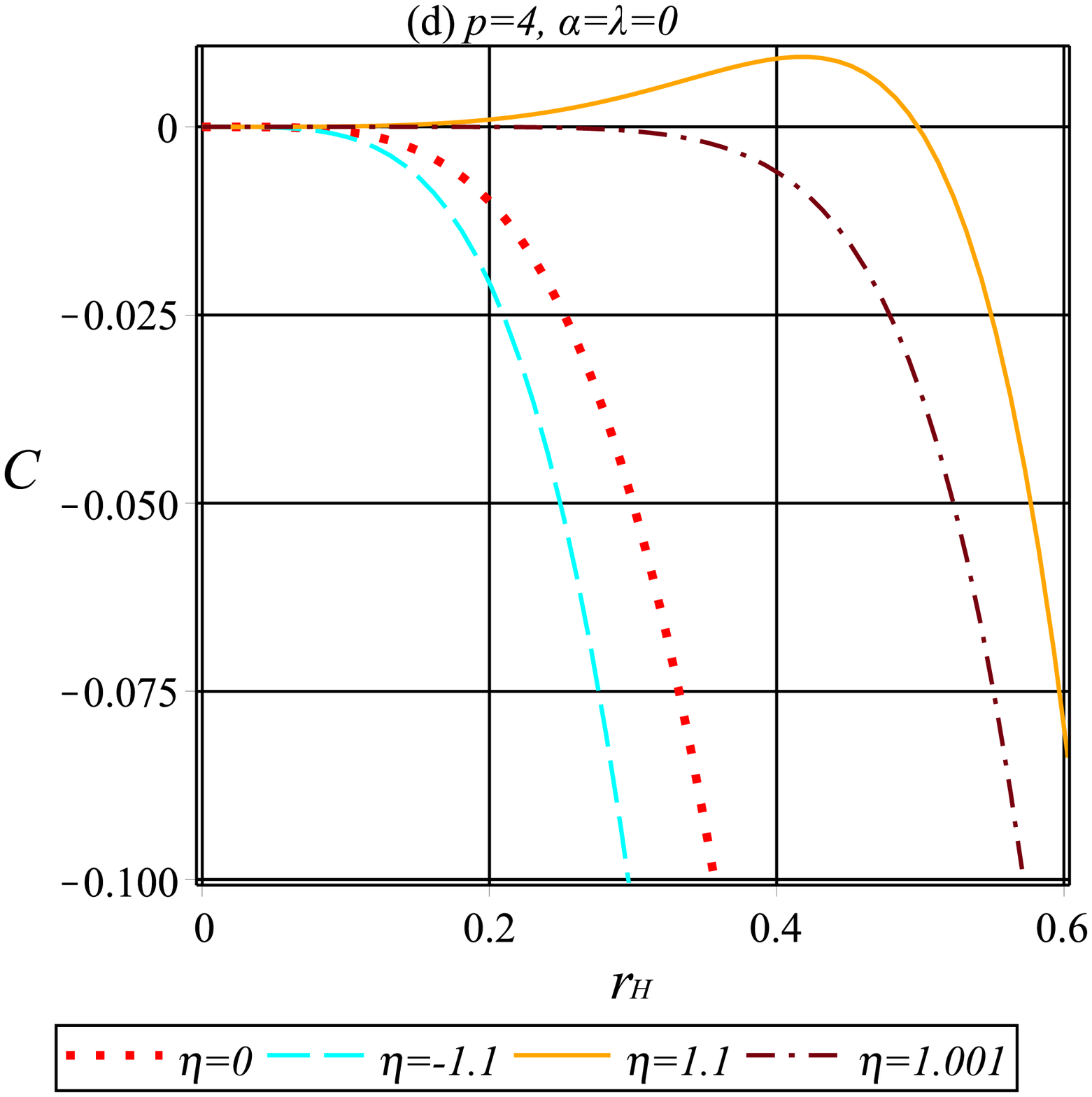}\includegraphics[width=50 mm]{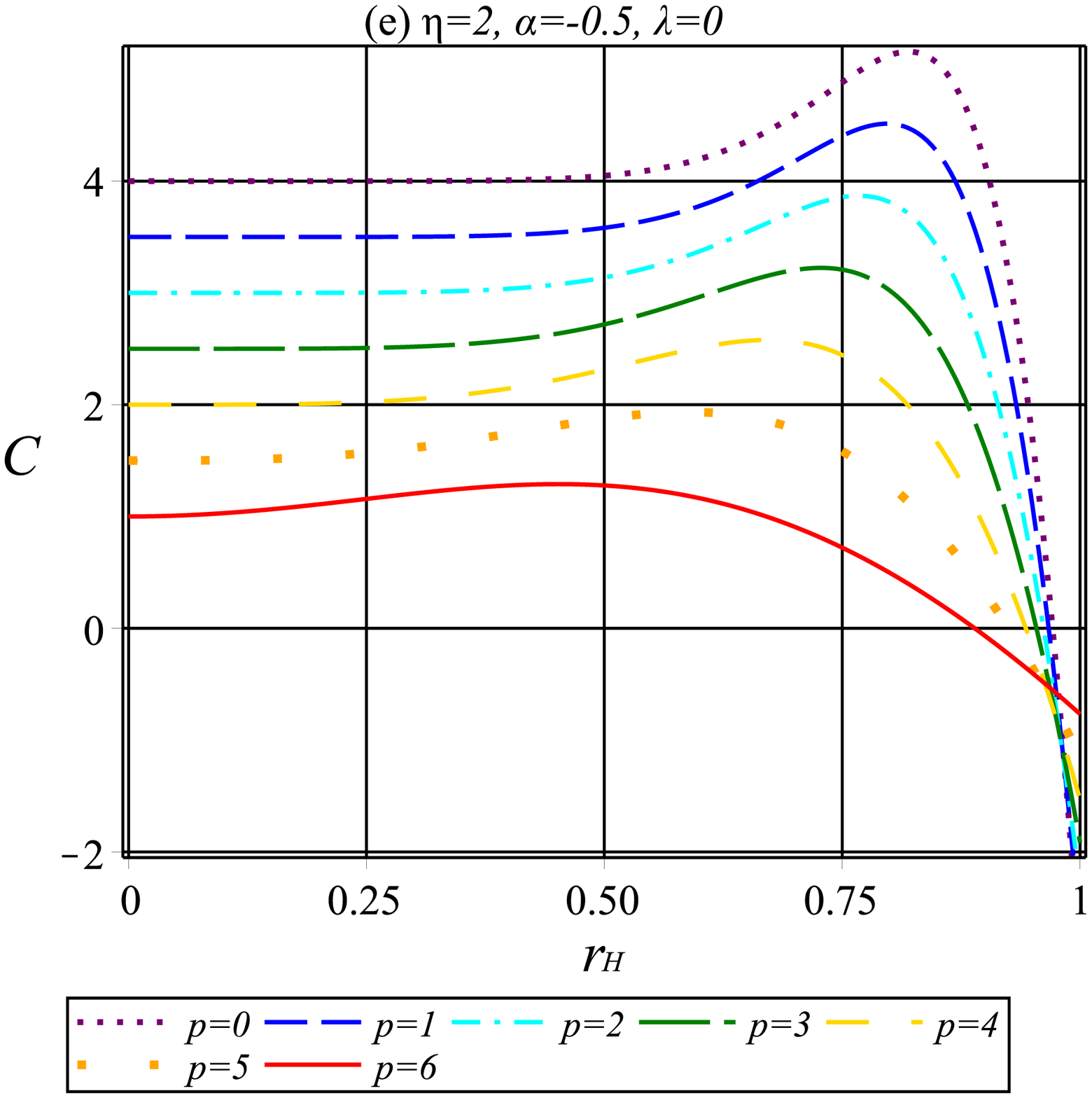}\includegraphics[width=50 mm]{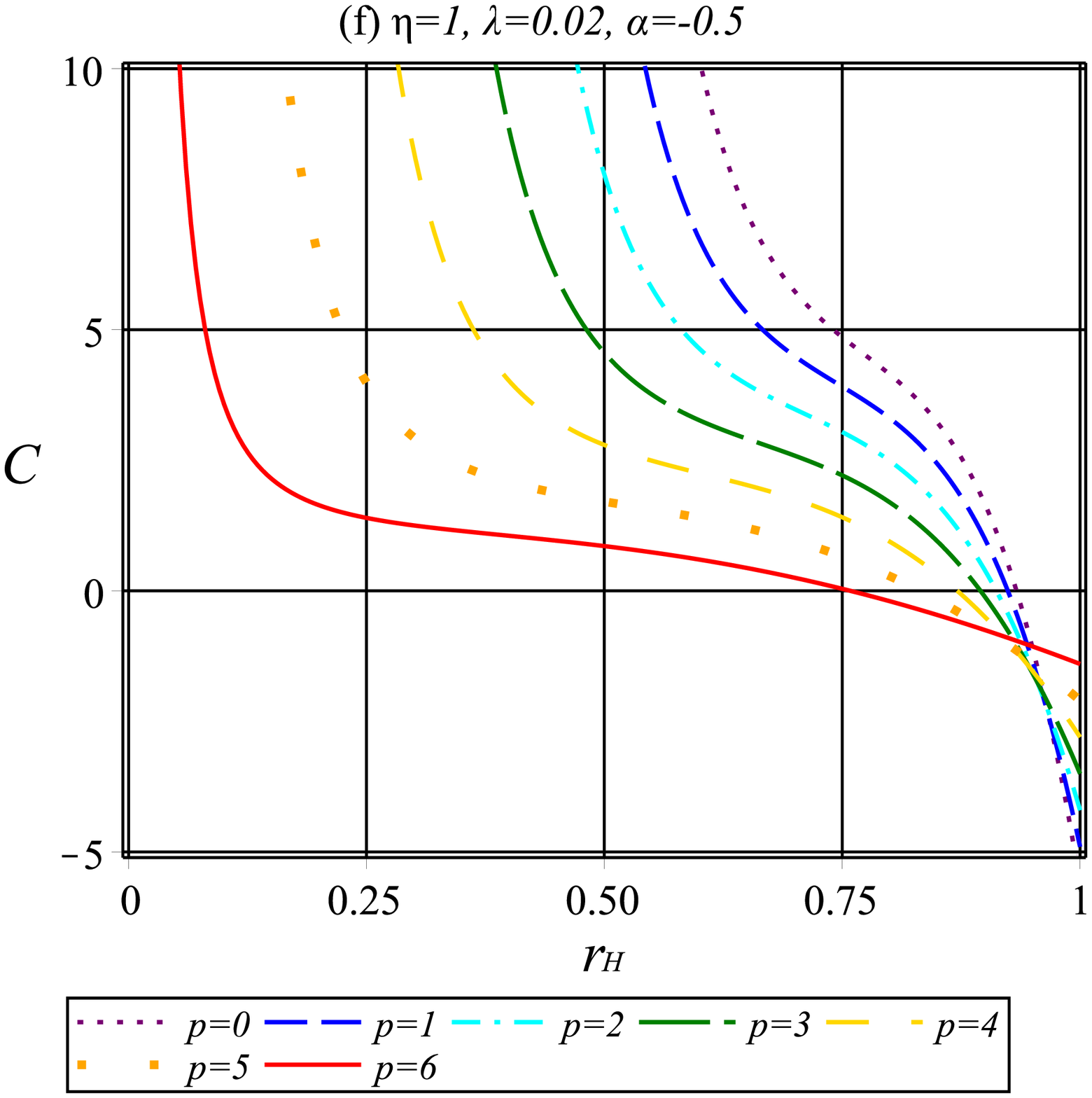}
\end{array}$
\end{center}
\caption{Specific heat in terms of horizon radius for $\beta=1$.}
\label{fig1}
\end{figure}

We note that the original entropy decreases by increasing temperature (decreasing radius). However, due to the perturbative logarithmic and non-perturbative exponential corrections, the value of corrected entropy becomes negative at a sufficiently small radius.   So,   the presence of the second order perturbative correction (with coefficient $\lambda$) is necessary to have positive entropy. This entropy increases as the radius of the horizon become small.
It is possible to calculate the effect of these perturbative and non-perturbative corrections to the specific heat of this system. Thus, we can write the corrected specific heat for this system as
\begin{equation}\label{C}
C=\frac{8-p}{\omega_{0}\cosh\beta r_{H}^{8-p}}\left[(\eta e^{-\omega_{0}\cosh\beta r_{H}^{8-p}}-1)(r_{H}^{8-p})^{2}\omega_{0}^{2}\cosh^{2}\beta-\alpha\omega_{0}\cosh\beta r_{H}^{8-p}+\lambda\right],
\end{equation}
where we have defined $\omega_{0}=\frac{\Omega_{8-p}}{4G_{10-p}}$. We can now   work in unit of $\omega_{0}$. It is clear that in absence   of perturbative or non-perturbative corrections ($\alpha=\eta=\lambda=0$),   the   original specific heat   $C_{0}$   is given by
\begin{equation}\label{C0}
C_{0}=T\frac{dS}{dT}=-\omega_{0}(8-p)r_{H}^{8-p}\cosh\beta.
\end{equation}
This original specific heat is negative, and thus the model is thermodynamically unstable. This hold, when the size of the black brane is large. However,   for black branes at the quantum scale, the perturbative and non-perturbative corrections cannot be neglected. Thus, for quantum black branes, we cannot neglect the effects coming from terms with coefficients $\alpha$, $\eta$, and $\eta$. In order to analyze the effect of these quantum correction terms, we plot Fig. \ref{fig1}. The behavior of the uncorrected specific heat (\ref{C0}) is   investigated in Fig. \ref{fig1} (a). It demonstrates that $D_{0}$-brane has a   large negative specific heat, and variation of specific heat for a $D_{6}$-brane is slower than other $D_{p}$-branes. Also, we can observe  that, at the small radius, there is the quantum corrections have similar effects for various  differences dimensions.\\

In Fig. \ref{fig1} (b), we observe the effect of the perturbative logarithmic corrections ($\alpha\neq0$, while $\eta=\lambda=0$). We use both the positive and negative values of  $\alpha$, and observe that the system is stable at a small radius for the negative $\alpha$. In Fig. \ref{fig1} (b) we plot the behavior of specific heat for a $D_{4}$-brane. Other $D_{p}$-branes will also have similar behavior as mentioned by Fig. \ref{fig1} (a). By suitable choice of the coefficient, these branes can be made thermodynamically stable at the quantum radius. For such stable systems, the specific heat stays constant even after increasing the temperature.\\
In that case the logarithmic corrected specific heat is,
\begin{equation}\label{C alpha}
C(\alpha)=(p-8)\left({\omega_{0}\cosh\beta r_{H}^{8-p}+\alpha}\right),
\end{equation}
We can see that the logarithmic correction will be negligible for the larger radius, so the  first order approximation is  dominates till 
\begin{equation}\label{dom-C alpha}
r_{H}^{8-p}<<\frac{\alpha}{\omega_{0}\cosh\beta}.
\end{equation}
For example, our selected values of parameter yields to the domination range at $r_{H}<0.75$.\\

In Fig. \ref{fig1} (c), we add the  effect of the second order perturbative   corrections ($\lambda\neq0$, $\alpha\neq0$, while $\eta=0$). Now for positive values of $\lambda$, the system becomes stable for branes with a small radius. However, for negative values of $\lambda$ the system remains unstable even at a small radius. So, for such systems, the specific heat increases due to an increase  in temperature.\\
In that case the second order corrected specific heat is,
\begin{equation}\label{C lambda}
C(\alpha)=(p-8)\frac{\left(\omega_{0}\cosh\beta r_{H}^{8-p}\right)^{2}-\lambda}{\omega_{0}\cosh\beta r_{H}^{8-p}},
\end{equation}
The second order term dominates till,
\begin{equation}\label{dom-C lambda}
r_{H}^{2(8-p)}<<\frac{\lambda}{(\omega_{0}\cosh\beta)^{2}}.
\end{equation}
For the  selected values of parameter, we observe that this corresponds to the second order term  dominating till  $r_{H}<0.35$,  which is a smaller range than the first order correction.\\

We analyze the effects of non-perturbative corrections from Fig. \ref{fig1} (d). This is done by only analyze the  effect of exponential corrected entropy on the system   ($\eta\neq0$ while $\alpha=\lambda=0$). If we neglect the first and second order perturbative corrections, can obtain thermodynamic stability at a small horizon radius with positive correction term, however negative one yields to an unstable system. Thus, by decreasing the radius of the horizon, specific heat becomes positive due to non-perturbative quantum corrections (with positive coefficient). Hence $D_{p}$-branes may be thermodynamically stable at a small radius at the quantum scale. Fig. \ref{fig1} (d) has been plotted for $p=4$.   One can  similarly analyze the behavior for other $D_{p}$-branes. Here, the specific heat behaves as the specific heat in a system with a Schottky-like anomaly.
In this case, the specific heat corrected by non-perturbative  corrections can be expressed as 
\begin{equation}\label{C eta}
C(\eta)=-(p-8)\left(\omega_{0}\cosh\beta r_{H}^{8-p}\right)\left(\eta\exp(\omega_{0}\cosh\beta r_{H}^{8-p})-1\right),
\end{equation}
hence, the non-perturbative term dominates till,
\begin{equation}\label{dom-C eta}
r_{H}^{(8-p)}<<\frac{\ln{\eta}}{(\omega_{0}\cosh\beta)}.
\end{equation}
For the  selected values of parameters ($\eta=1.001$, $\omega_{0}=\beta=1$), we observe that this corresponds to the second order term  dominating till  $r_{H}<0.16$,  which is a smaller  than the scale at which  first and second order perturbative corrections dominate.\\

Here, the non-perturbative corrections  dominant at much   smaller scales, than the perturbative  cases.  Now even though this is expected from the general structure of the perturbation theory, the non-pertubative corrections dominate at small radius. Thus, they can change the thermodynamic stability of  $D_{p}$-branes at quantum scales. This can have important consequences in physical models for $D_{p}$-branes. 

Fig. \ref{fig1} (e),   shows the effects of logarithmic and exponential corrections simultaneously. Using some specific values of coefficients, we can observe that $D_{6}$-brane is unstable even for a small radius. However,   other configurations are stable at a small radius due to quantum corrections. The system again resembles a system with a   Schottky-like anomaly. The specific heat for this system becomes constant at high temperatures.
Finally, in Fig. \ref{fig1} (f), we can see the combined  effects of both first and second order perturbative corrections along with the exponential non-perturbative  correction.   We observe that the $D_{p}$-brane configuration is stable at a critical value of the radius of the horizon, and the specific heat increases with  temperature.\\

It may be interesting to compare the second order correction with the non-perturbative one which yields to the following relation,
\begin{equation}\label{C eta lambda}
\frac{\eta}{\lambda}=\frac{\exp(\omega_{0}\cosh\beta r_{H}^{8-p})}{\left(\omega_{0}\cosh\beta r_{H}^{8-p}\right)^{2}},
\end{equation}
which clearly demonstrates  that at small radius the non-perturbative corrections dominate.\\

\begin{figure}[h!]
\begin{center}$
\begin{array}{cccc}
\includegraphics[width=70 mm]{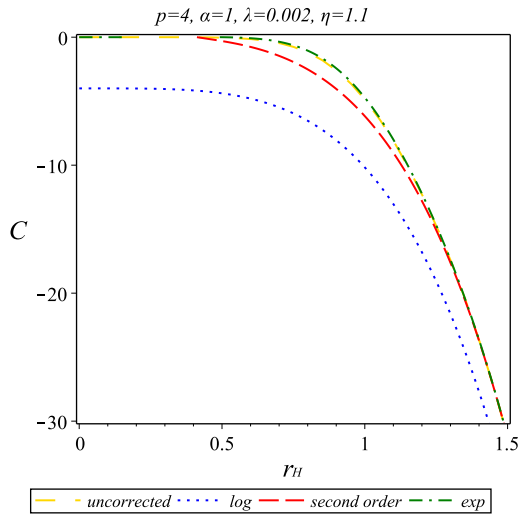}
\end{array}$
\end{center}
\caption{Specific heat in terms of horizon radius for $\beta=1$ and $\omega_{0}=1$.}
\label{figC}
\end{figure}

In order to compare the effects of correction terms we represent Fig. \ref{figC}. We can see that non-perturbative contribution decays at larger radius, and can be neglected for large enough radius.

Now, we can obtain the Helmholtz free energy via the following general formula,
\begin{equation}\label{F-general}
F=-\int{SdT},
\end{equation}
where we can use $dT=\frac{dT}{dr_{H}}dr_{H}$ to solve the integral. Using the temperature (\ref{T}) and entropy (\ref{S}) we find,
\begin{equation}\label{F}
F=\frac{\omega_{0}r_{H}^{7-p}}{4\pi}+F_{\alpha}+F_{\lambda}+F_{\eta},
\end{equation}
where we defined,
\begin{eqnarray}\label{F alpha}
F_{\alpha}&\equiv&-\frac{(7-p)\alpha}{4\pi r_{H}\cosh\beta}\left(\ln{(\omega_{0}r_{H}^{8-p}\cosh\beta)}+8-p\right),\nonumber\\
F_{\lambda}&\equiv&-\frac{\lambda(7-p)}{4\pi(9-p)\omega_{0}r_{H}^{9-p} \cosh^{2}\beta},\nonumber\\
F_{\eta}&\equiv&\left(\frac{\eta(\omega_{0}\cosh\beta)^{-\frac{7-p}{2(8-p)}-2}\omega_{0}}{4\pi(2p-15)}\right)e^{-\frac{\omega_{0}\cosh\beta}{2}r_{H}^{8-p}}
r_{H}^{-\frac{3p^{2}-49p+200}{2(8-p)}}X,
\end{eqnarray}

here, $X$ is given in terms of Whittaker function as,
\begin{eqnarray}\label{X}
X=&-&(8-p)\left((8-p)\omega_{0}\cosh\beta r_{H}^{8-p}-(7-p)\right)WM\left(-\frac{9-p}{2(8-p)},-\frac{2p-15}{2(8-p)},\omega_{0}\cosh\beta r_{H}^{8-p}\right)\nonumber\\
&+&(7-p)^{2}WM\left(\frac{7-p}{2(8-p)},-\frac{2p-15}{2(8-p)},\omega_{0}\cosh\beta r_{H}^{8-p}\right),
\end{eqnarray}

\begin{figure}[h!]
\begin{center}$
\begin{array}{cccc}
\includegraphics[width=50 mm]{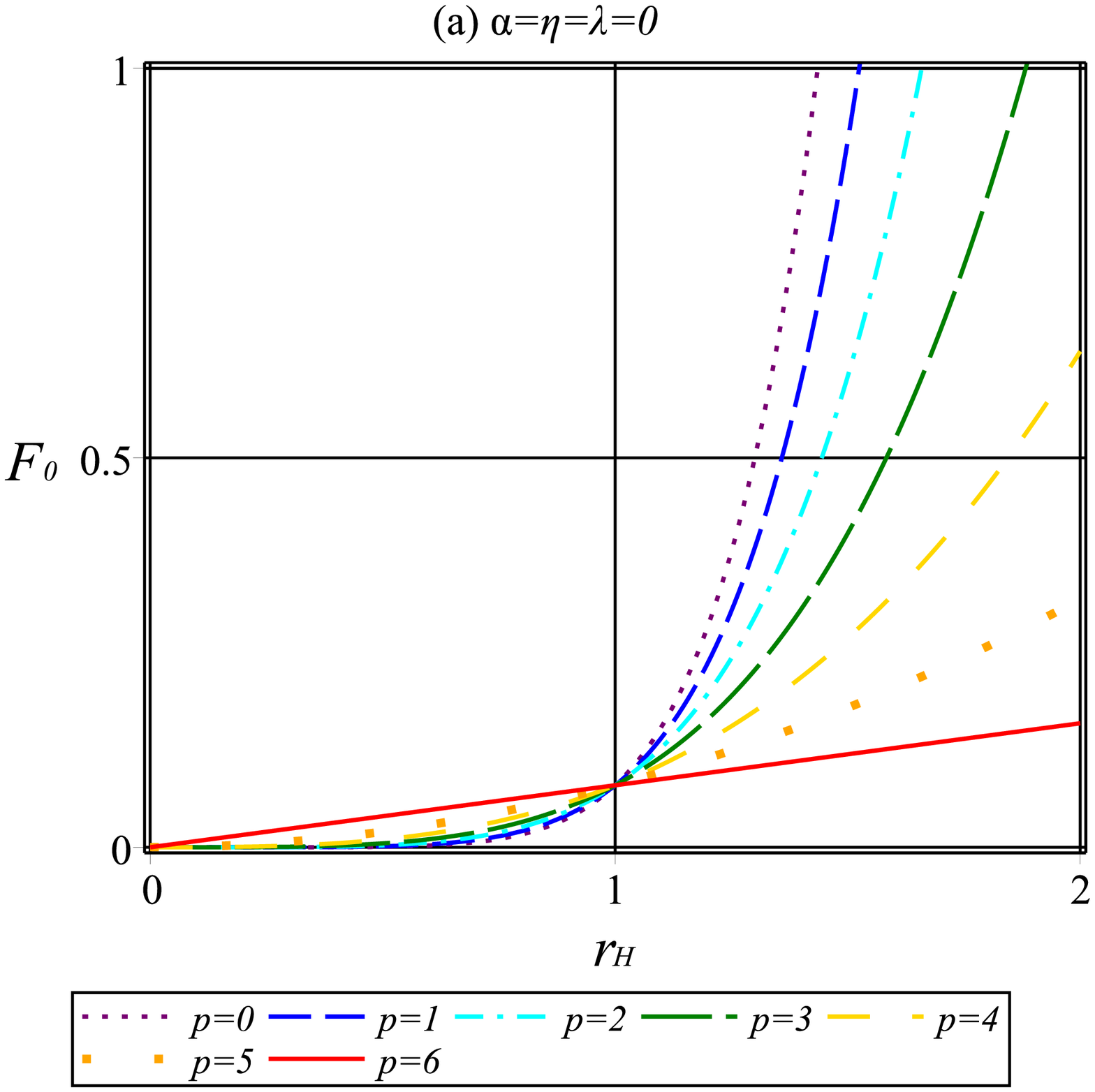}\includegraphics[width=50 mm]{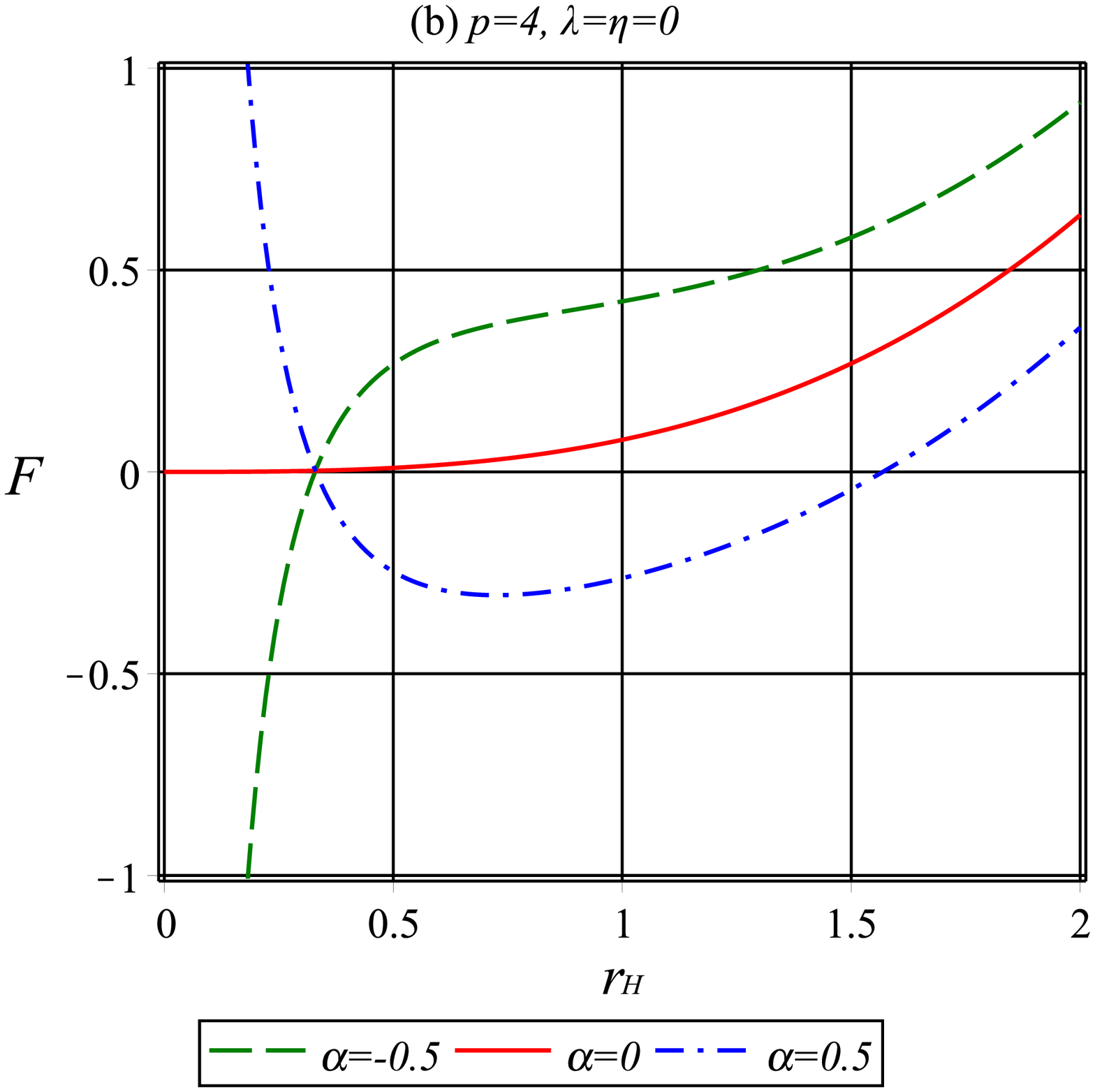}\includegraphics[width=50 mm]{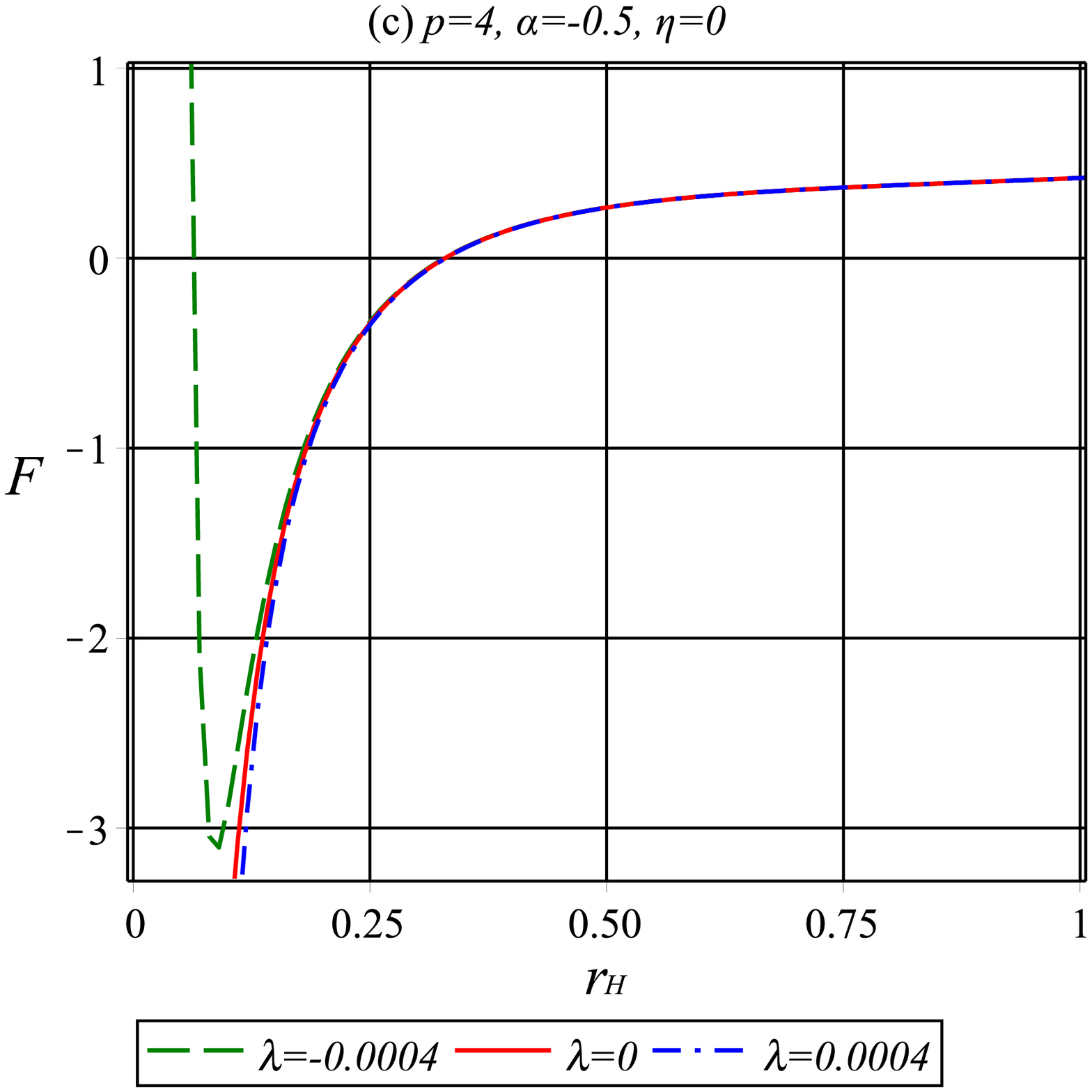}\\
\includegraphics[width=50 mm]{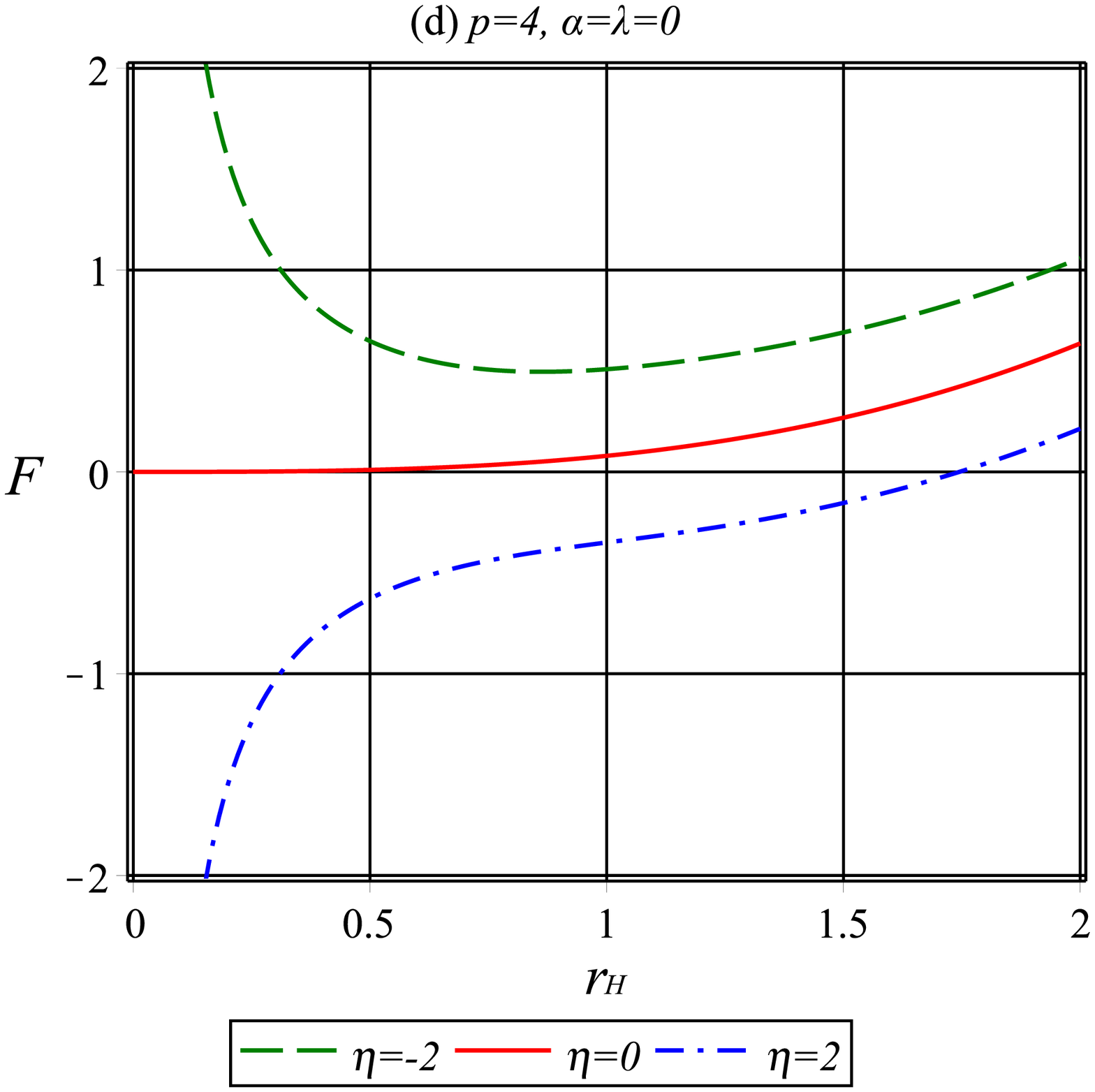}\includegraphics[width=50 mm]{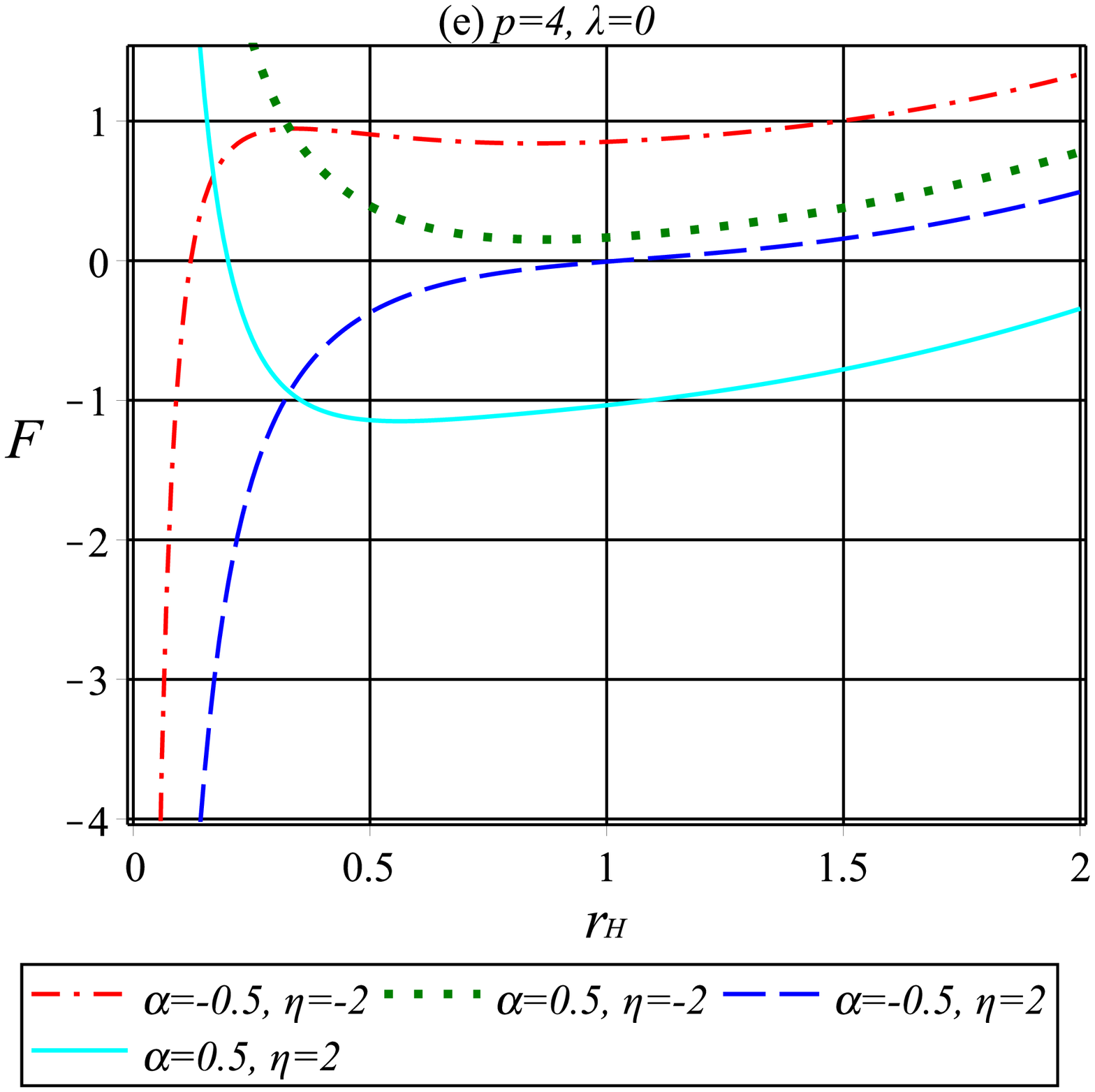}\includegraphics[width=50 mm]{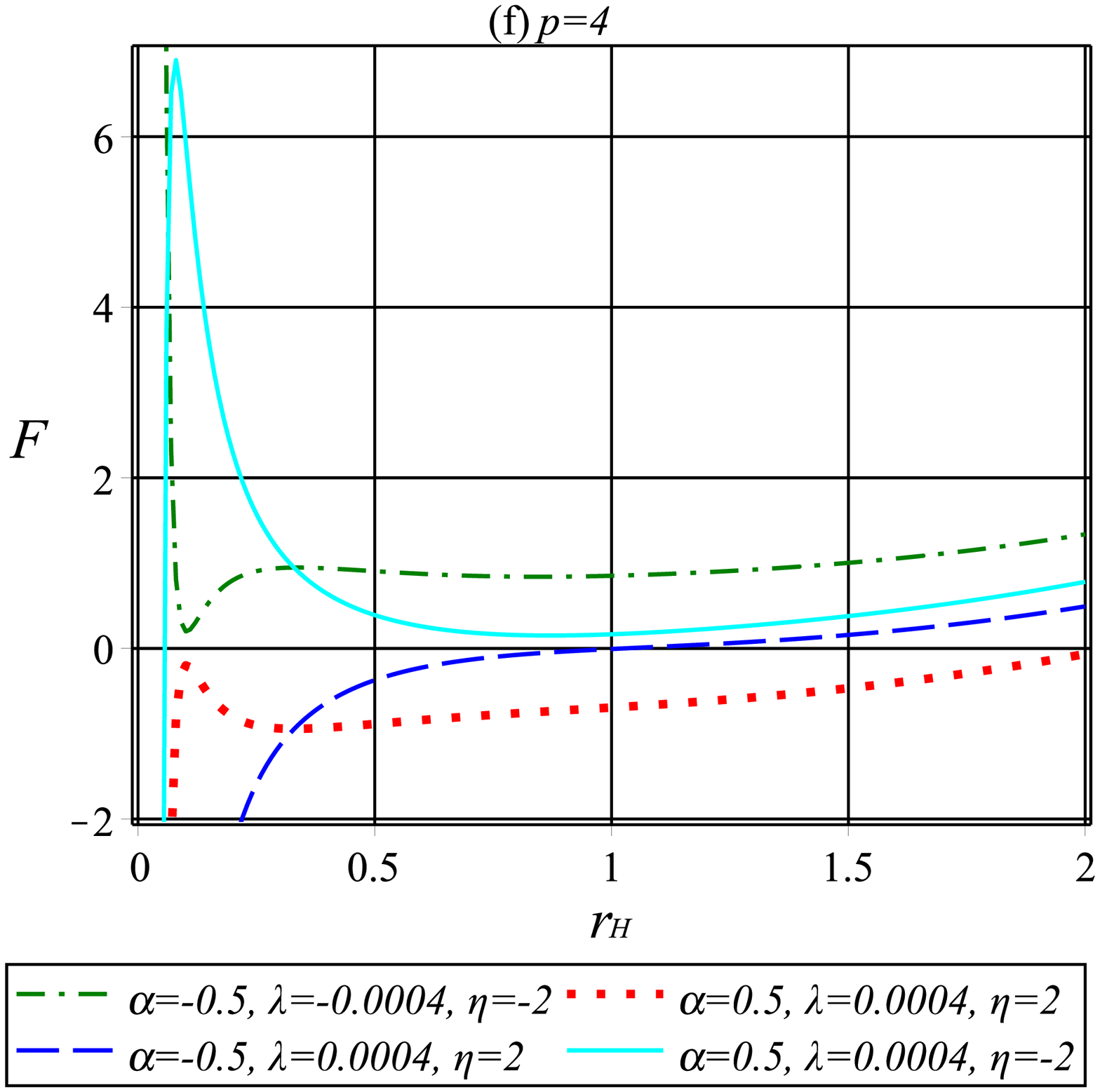}
\end{array}$
\end{center}
\caption{Helmholtz free energy in terms of horizon radius for $\beta=1$.}
\label{fig2}
\end{figure}

As we can see from Fig. \ref{fig2} (a), Helmholtz free energy is positive in absence of quantum corrections. It may be observed from a solid red line of Fig. \ref{fig2} (a), that the Helmholtz free energy of $D_{6}$-brane is linear for $r_{H}$. There is a special radius ($r_{H}=1$), where Helmholtz free energy of all $D_{p}$-branes are the same. For   $D_p$ branes, the   Helmholtz free energy decreases with temperature. Effect of the logarithmic correction on the Helmholtz free energy illustrated in Fig. \ref{fig2} (b).  As expected, we can see that leading order correction is important at small horizon radius. As we found in Fig. \ref{fig1} (b), stability at small radii obtained by negative logarithmic correction coefficient, which is corresponding to the negative Helmoltz free energy represented by green dashed line of Fig. \ref{fig2} (b).\\
Then, in Fig. \ref{fig2} (c), we can analyze the effects of the second order perturbative correction. We can see that at smaller horizon radius the second order correction with positive coefficient is dominant and may yields to instability of our system which is coincident with result of Fig. \ref{fig1} (c).\\
In Fig. \ref{fig1} (d), we show effect of non-perturbative correction for the case of $p=4$ (other dimensions yields to the similar result). Fig. \ref{fig2} (e) show the effect of both the first order perturbative and non-perturbative corrections. We can see that both stability and instability at small radii can reach depends on the sign of coefficients. Finally, in Fig. \ref{fig2} (f) we can see effect of all correction simultaneously. We can see that there are some situations of complete stability or instability at small radii depends on the correction coefficient sign.

\section{Near-Extremal Solution}
Now, we can analyze the effect of perturbative and non-perturbative quantum corrections on black branes near extremality.
For such black branes, it has been observed that $r_H$ is proportional to the energy density above extremality    \cite{Peet:2000hn},
\begin{equation}\label{F eta}
r_{H}\propto\varepsilon,
\end{equation}
where constant of proportionality is $8\sqrt{\pi^{p-7}}G_{10}\Gamma(\frac{7-p}{2})$. So, we can write the   temperature for such a system as
\begin{equation}\label{T ex}
T\sim c_{T}\varepsilon^{\frac{5-p}{2(7-p)}}
\end{equation}
where $c_{T}$ is constant. Now, the corrected entropy for this system can be written as
\begin{equation}\label{S ex}
S_{BH}\sim c_{S}\varepsilon^{\frac{9-p}{2(7-p)}}+\alpha\ln{c_{S}\varepsilon^{\frac{9-p}{2(7-p)}}}
+\frac{\lambda}{\varepsilon^{\frac{9-p}{2(7-p)}}}+\eta e^{-c_{S}\varepsilon^{\frac{9-p}{2(7-p)}}},
\end{equation}
where $\lambda=\frac{\gamma}{4c_{S}}$. The specific heat of near-extremal case can now be written as,
\begin{equation}\label{C ex}
C=\frac{9-p}{5-p}\frac{c_{S}\left(1-\eta e^{-c_{S}\varepsilon^{\frac{9-p}{2(7-p)}}}\right)\varepsilon^{\frac{9-p}{7-p}}+\alpha\varepsilon^{\frac{9-p}{2(7-p)}}-\lambda}{\varepsilon^{\frac{9-p}{2(7-p)}}}.
\end{equation}
We observe that the near-extremal $D_{6}$-brane is unstable in absence of perturbative quantum corrections. However, exponential and second order corrections can make this system stable at small quantum radius. Situation is different   for the $D_{p}$-branes with $p\leq4$.
We can obtain the corrections to the internal energy of this system as
\begin{equation}\label{E}
E=\frac{c_{0}(9-p)\varepsilon}{2(7-p)}+E_{\alpha}+E_{\lambda}+E_{\eta},
\end{equation}

\begin{figure}[h!]
\begin{center}$
\begin{array}{cccc}
\includegraphics[width=50 mm]{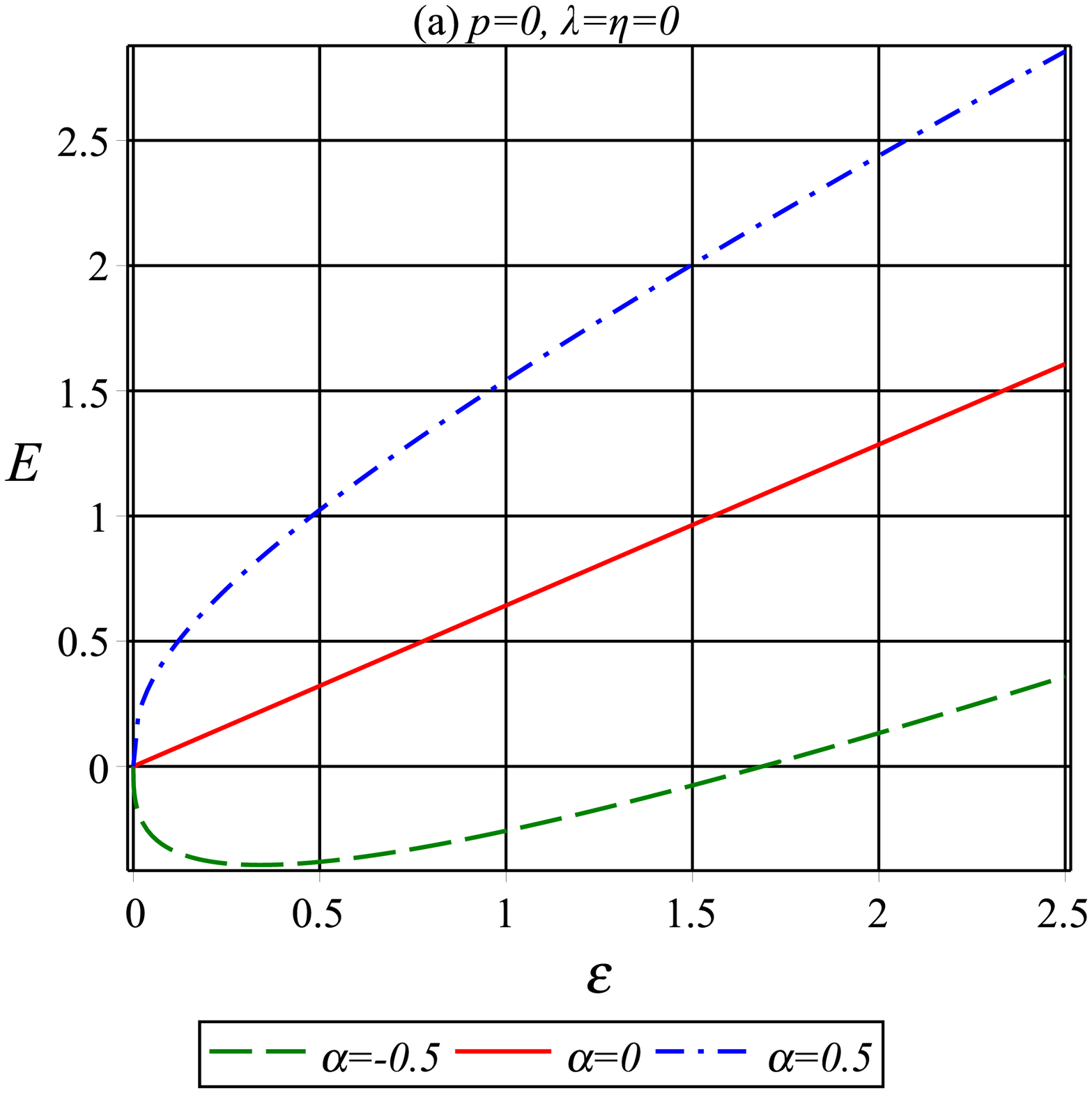}\includegraphics[width=50 mm]{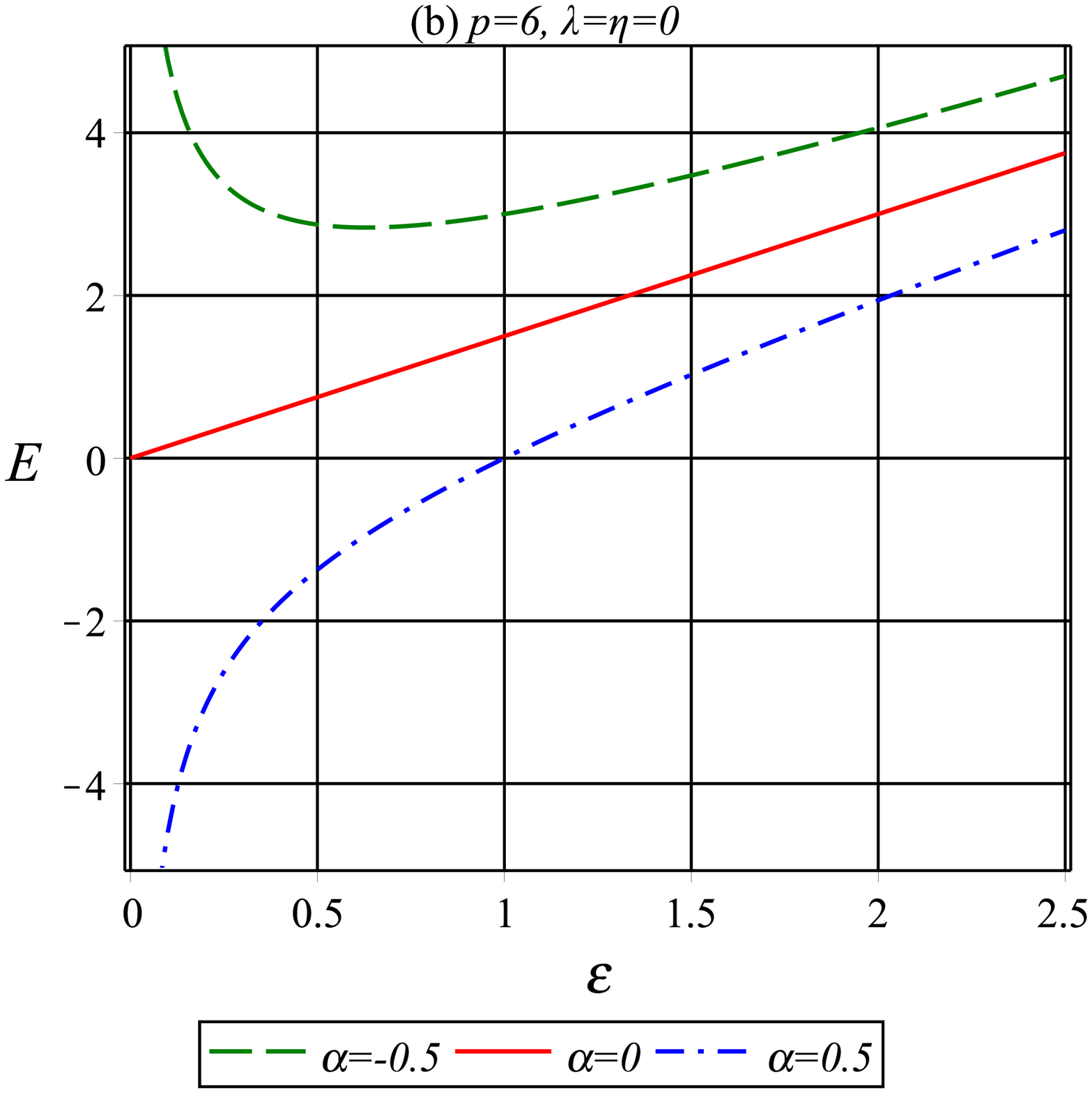}
\end{array}$
\end{center}
\caption{Effect of logarithmic correction on the internal energy for $c_{T}=c_{S}=1$.}
\label{fig3}
\end{figure}

where $c_{0}=c_{T}c_{S}$, and   we   have also defined,
\begin{eqnarray}\label{E alpha}
E_{\alpha}&=&\frac{\alpha(9-p)}{5-p}c_{T}\varepsilon^{\frac{5-p}{2(7-p)}},
\nonumber \\
E_{\lambda}&=&\lambda\frac{c_{T}}{4}(9-p)\varepsilon^{-\frac{2}{7-p}},
\nonumber \\
E_{\eta}&=&\frac{c_{1}\eta\varepsilon^{-\frac{9-p}{2(7-p)}}(9-p)}{12(7-p)(5-p)(8-p)(23-3p)}e^{-\frac{c_{S}}{2}e^{\frac{9-p}{2(7-p)}}}
\varepsilon^{-\frac{p^{3}-21p^{2}+143p-315}{4(9-p)(7-p)^{2}}}U,
\end{eqnarray}
with $c_{1}=c_{T}c_{S}^{-\frac{23-3p}{2(9-p)}}$.\\

\begin{figure}[h!]
\begin{center}$
\begin{array}{cccc}
\includegraphics[width=50 mm]{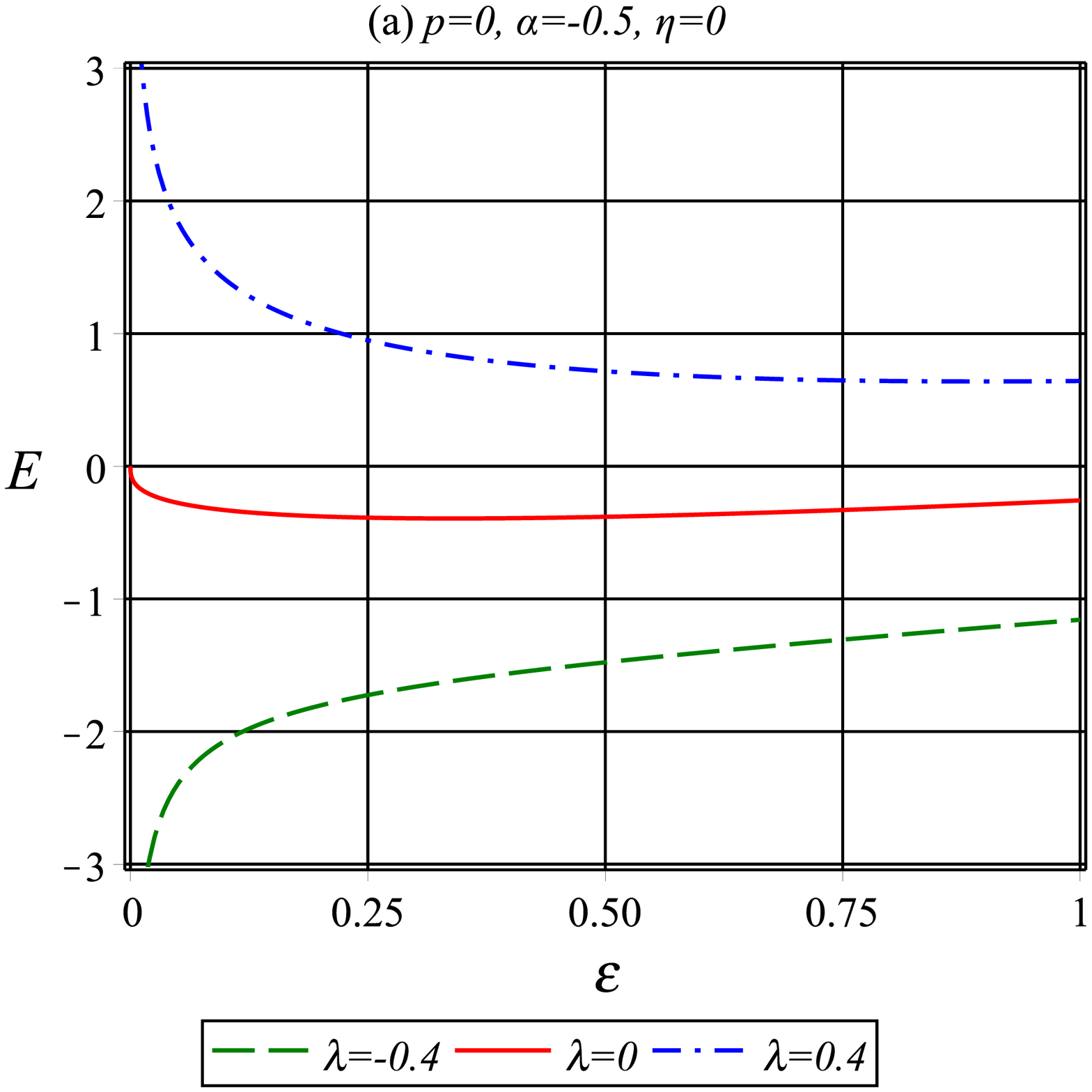}\includegraphics[width=50 mm]{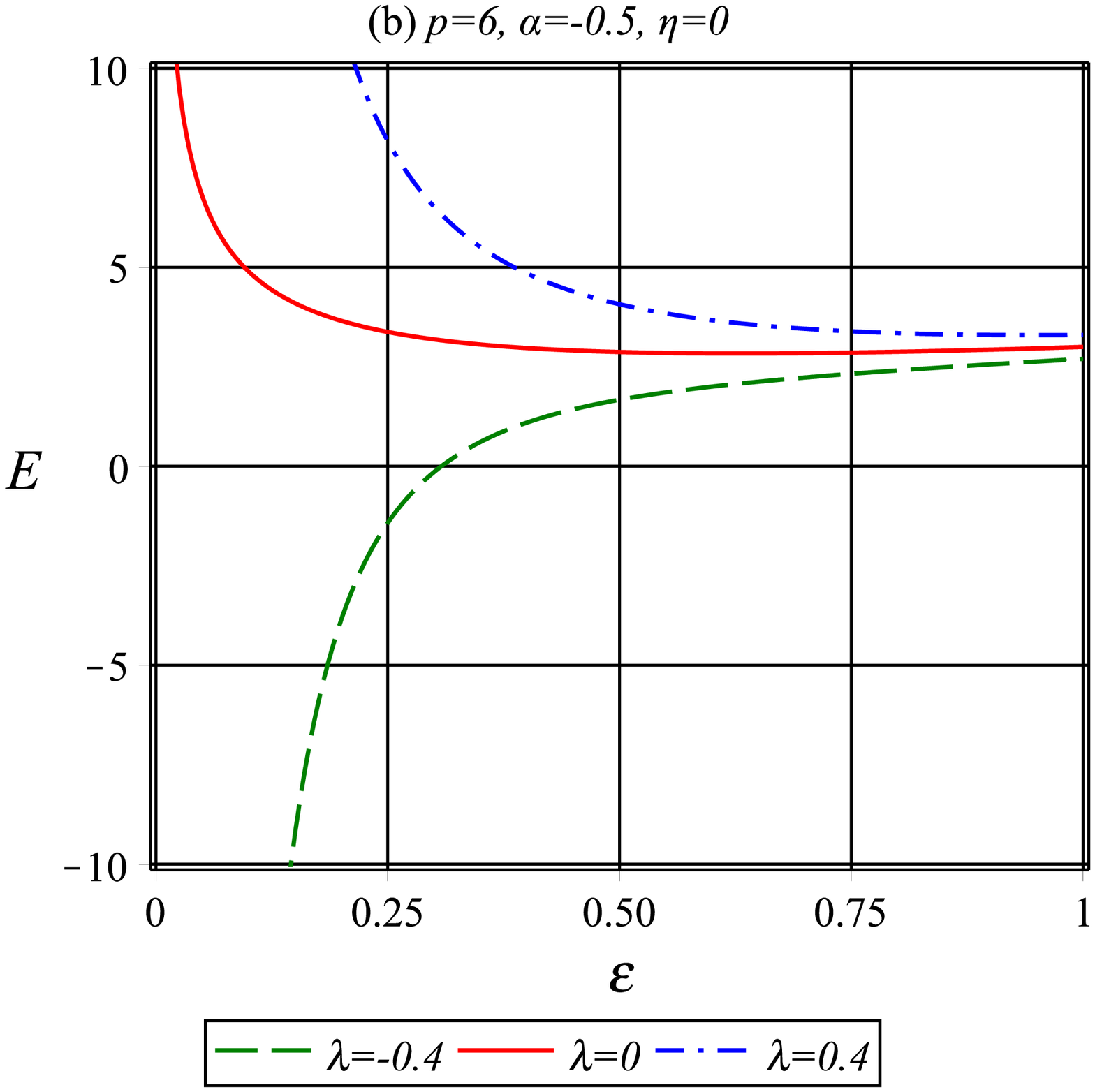}
\end{array}$
\end{center}
\caption{Effect of perturbative corrections on the internal energy for $c_{T}=c_{S}=1$.}
\label{fig4}
\end{figure}

In the last expression of (\ref{E alpha}), $U$ is given in terms of Whittaker function as,
\begin{eqnarray}\label{U}
U\equiv&-&\left(c_{S}(9-p)\varepsilon^{\frac{9-p}{2(7-p)}}+23-3p\right)(9-p)WM\left(\frac{5-p}{2(9-p)},\frac{2(8-p)}{9-p},c_{S}\varepsilon^{\frac{9-p}{2(7-p)}}\right)\nonumber\\
&+&(23-3p)^{2}WM\left(\frac{23-3p}{2(9-p)},\frac{2(8-p)}{9-p},c_{S}\varepsilon^{\frac{9-p}{2(7-p)}}\right).
\end{eqnarray}
In the absence of correction terms, the internal energy is linear for $\varepsilon$. This original internal energy is denoted by $E_{0}$. We find that quantum corrections are important for small $\varepsilon$, from plots of Fig. \ref{fig3}, as they can change the value of the internal energy. It is also observed that the situation may be different for $p\leq4$ and $p=6$. Hence, we plotted two cases of $p=0$ in Fig. \ref{fig3} (a), and $p=6$ in Fig. \ref{fig3} (b), to investigate   the effects of the logarithmic correction term.\\
In plots of Fig. \ref{fig4} we can see effects of complete perturbative corrections. As expected, we show  that the second order correction is dominant at smaller $\varepsilon$.\\

\begin{figure}[h!]
\begin{center}$
\begin{array}{cccc}
\includegraphics[width=50 mm]{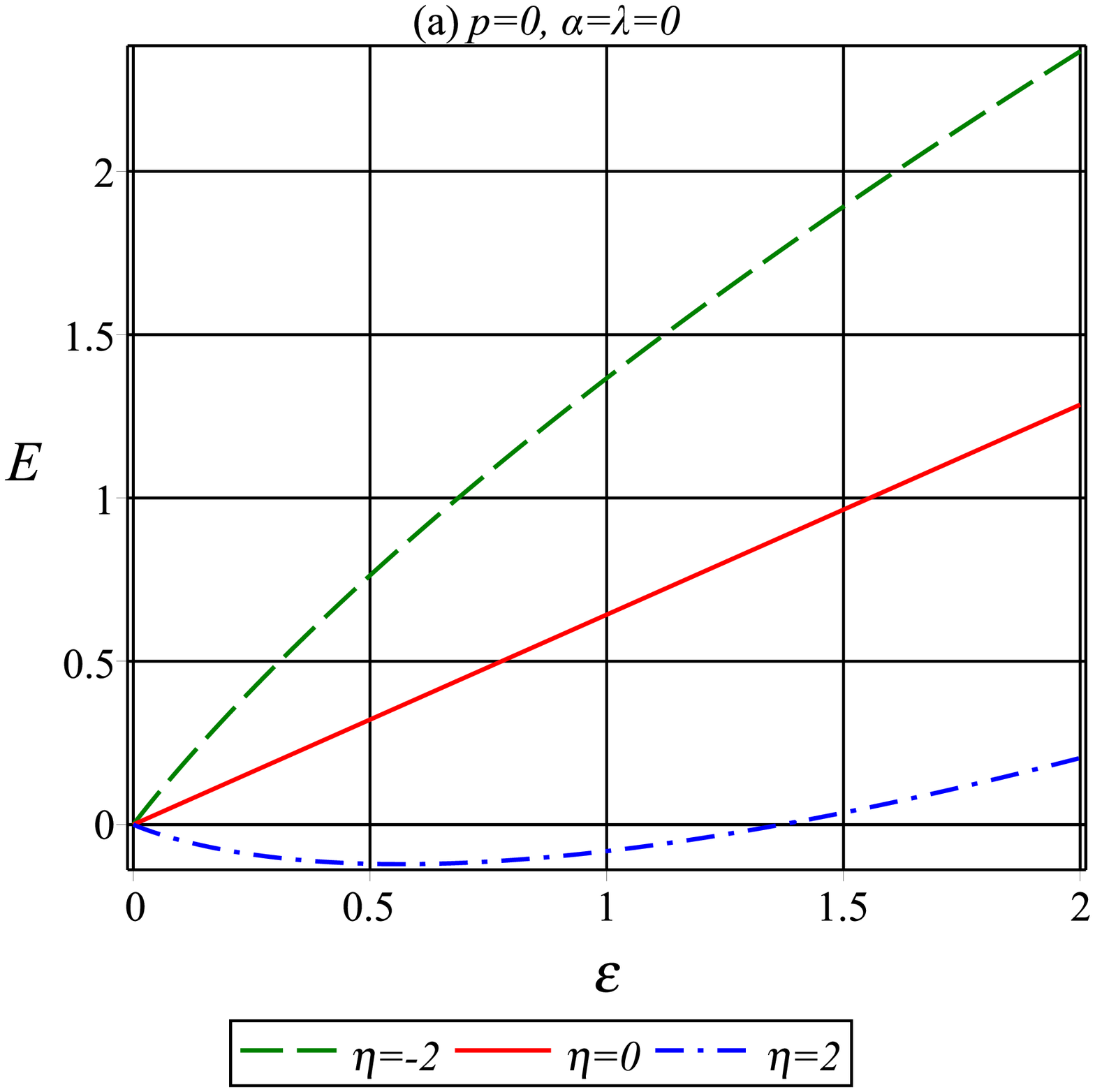}\includegraphics[width=50 mm]{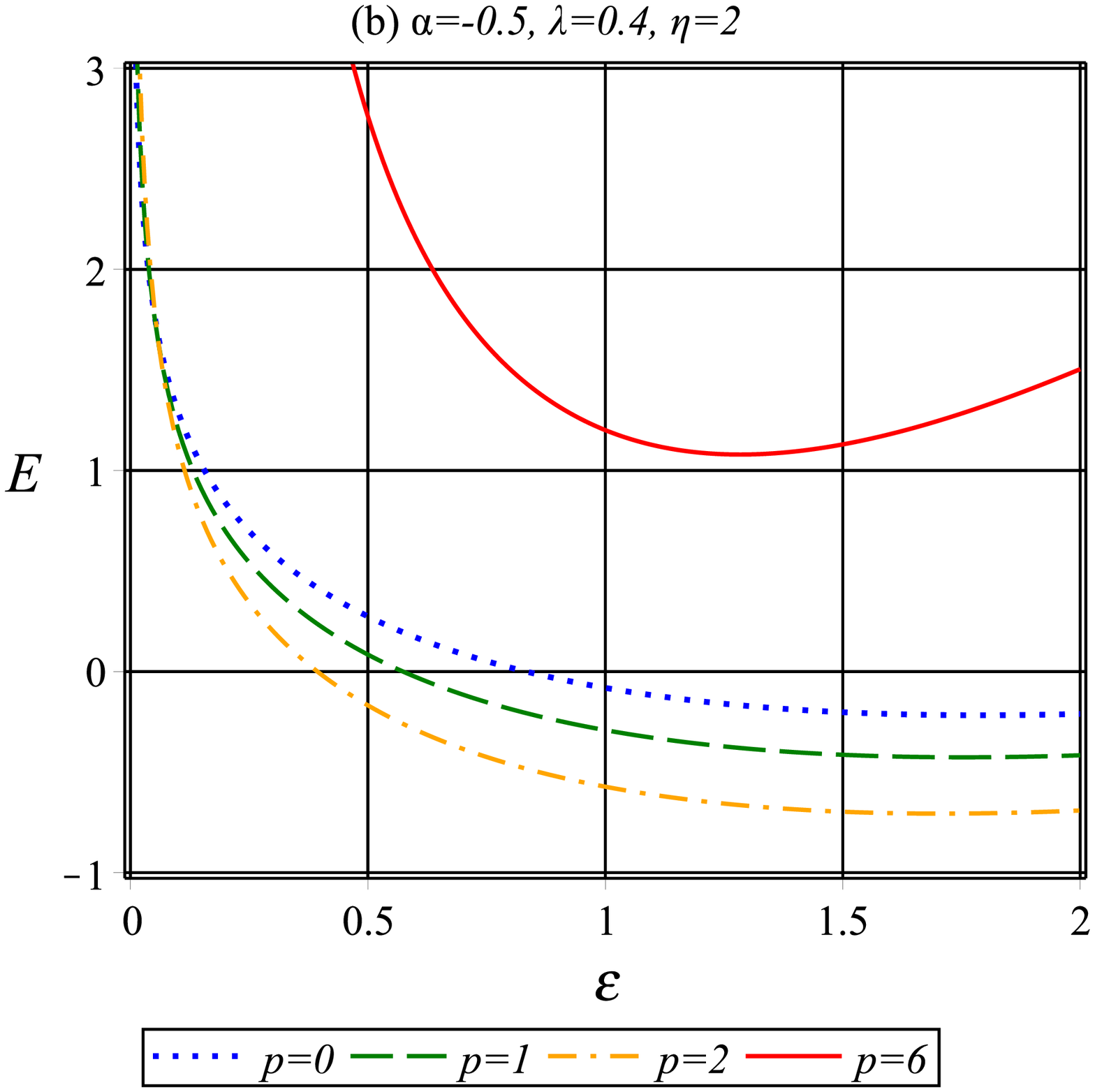}
\end{array}$
\end{center}
\caption{Effect of non-perturbative corrections on the internal energy for $c_{T}=c_{S}=1$. Cyan lines are corresponding to $\alpha=\lambda=\eta=0$ for $p=0$ (dotted) and $p=6$ (solid).}
\label{fig5}
\end{figure}

Then, from Fig. \ref{fig5} (a), we show effect of non-perturbative correction. we can observe the effects of all corrections together in Fig. \ref{fig5} (b),   and compare them to the system without such corrections (cyan lines of Fig. \ref{fig5} (a)).
The logarithmic corrections increase the value of the internal energy of $D_{6}$-brane, while other corrections decrease its value. Due to the combined effect of both perturbative and non-perturbative corrections, there is a minimum value of the internal energy at high temperatures.\\

Now, we can obtain the corrected   Helmholtz free energy for this system as
\begin{equation}\label{F-ext}
F=-\frac{c_{0}(5-p)\varepsilon}{2(7-p)}+F_{\alpha}+F_{\lambda}+F_{\eta},
\end{equation}
where we have defined,
\begin{eqnarray}\label{F-ext alpha}
F_{\alpha}&\equiv&\frac{\alpha}{5-p}c_{T}e^{\frac{5-p}{2(7-p)}\ln{\varepsilon}}\left[9-p-(5-p)\ln{(c_{S}\varepsilon e^{-\frac{(5-p)\ln{\varepsilon}}{2(7-p)}})}\right],
\nonumber \\
F_{\lambda}&\equiv&\lambda\frac{c_{T}}{4}(5-p)\varepsilon^{-\frac{2}{7-p}},
\nonumber \\
F_{\eta}&\equiv&-\frac{c_{2}\eta Y}{6p^{2}-88p+322}\varepsilon^\frac{p^{3}-27p^{2}+239p-693}{(9-p)(7-p)^2}e^{-\frac{c_{S}}{2}e^{\frac{9-p}{2(7-p)}}},
\end{eqnarray}
with $c_{2}=c_{T}c_{S}^{-\frac{2p^{2}+30p-112}{(9-p)(7-p)}}$. Here, $Y$ is given in terms of Whittaker function as,
\begin{eqnarray}\label{Y}
Y=&-&\left(c_{S}(9-p)\varepsilon^{\frac{9-p}{2(7-p)}}-2(7-p)\right)(9-p)WM\left(-\frac{2}{9-p},\frac{23-3p}{2(9-p)},c_{S}\varepsilon^{\frac{9-p}{2(7-p)}}\right)\nonumber\\
&+&4(7-p)^{2}WM\left(\frac{7-p}{9-p},\frac{23-3p}{2(7-p)}, c_{S}\varepsilon^{\frac{9-p}{2(7-p)}}\right).
\end{eqnarray}
Now, if we neglect the perturbative corrections, the Helmholtz free energy of $D_{6}$-brane will become negative at small $\varepsilon$. It will also become negative for $D_{p}$-brane, with $p\leq4$   at small $\varepsilon$,   even after quantum corrections. So, the quantum corrections can have important consequences for the behavior of the Helmholtz free energy of $D_p$ branes.

\section{Quantum Corrected Geometry}
It may be noted that geometry emerges from thermodynamics in the Jacobson formalism  \cite{gr12}. So, the thermal fluctuations to the geometry of black branes occur due to quantum fluctuations of its geometry. So,    such quantum corrections to the geometry of black holes can be explicitly constructed from their thermal fluctuations  \cite{gr14, gr16}.
Now in Jacobson formalism \cite{gr12},
 one can choose a two-surface element $\cal P$
any point $p$. Here a Killing field $\chi^a$ generates  orthogonal boosts   such
that the temperature $T$ is the Unruh temperature. This Unruh temperature can be expressed as  $T =\hbar \kappa/2\pi$ with the acceleration of the Killing orbit represented by $\kappa$. The boost-energy current $T_{ab}\chi^a$ is used to obtain the heat flow. We also consider a local Rindler horizon through $p$. It is  generated
by $\chi^a$ and its  future points to the energy carried by matter. It may be noted that the
  past pointing heat flux through $\cal P$,
can be denoted by $\cal H$.  Now we can write
\begin{equation}
\delta Q=\int_{\cal H}T_{ab} \chi^a d\Sigma ^b, \label{1J}
\end{equation}
Here $k^a$ a tangent vector to the horizon, and  $d\Sigma ^a = k^a d\lambda d\cal A$.
 The  affine parameter $\lambda$ vanishing at $\cal P$, with negative values to
the past of $\cal P$. The area
element is denoted by  $d\cal A$, and using this area element, we can write
\begin{equation}
\delta Q=-\kappa\int_{\cal H}\lambda T_{ab} k^a k^b d\lambda d\cal A, \label{2J}
\end{equation}
As the entropy  is assumed to be proportional to the horizon area.  Using  the  expansion of the horizon generated
by $\theta$,  we obtain
\begin{equation}
\delta {\cal A}=-\kappa\int_{\cal H}\theta d\lambda d\cal A, \label{3J}
\end{equation}
The   Einstein equations can be obtained by  expanding the $\theta$ terms. We note that the
 Raychaudhuri equation, they    vanish at $\cal P$ by a suitable choice of the local Rindler horizon
\begin{equation}
\frac{d\theta}{d\lambda}=-\frac{1}{2}\theta^2-\sigma^2-R_{ab}k^ak^b. \label{4J}
\end{equation}
Now   integrate this equation,  we can write
\begin{equation}
\delta {\cal A}=-\int_{\cal H}\lambda R_{ab} k^a k^b d\lambda d\cal A, \label{5J}
\end{equation}
Thus, we observe that
 $\delta Q = TdS = \left(\hbar\kappa/2\pi\right)\xi\delta\cal A$
holds, if   $T_{ab}k^ak^b =  \left(\hbar\xi/2\pi\right) R_{ab}k^ak^b$
for all null $k^a$. Thus, we obtain   $\left(2\pi/\hbar\xi\right)T_{ab} = R_{ab} +fg_{ab}$ for a suitable  function $f$.
Now using conservation laws and the Bianchi, we can write
$f = -R/2+\Lambda$ for some constant
$\Lambda$. So, we obtain   Einstein equations from thermodynamics of this system
\begin{equation}
R_{ab}-\frac{1}{2}+\Lambda g_{ab}=\frac{2\pi}{\hbar\xi}T_{ab}
\end{equation}
Here  $\xi$  is related  to Newton constant as
$G = \left(4\hbar\xi \right)^{-1}$.
Thus, we
will try to  obtain modified metric, such that the modified metric  reproduce the corrected entropy obtained in  Eq. (\ref{S}). We first note that the modified metric for black branes can be written as
\begin{equation}\label{metric2}
ds^{2}=\frac{1}{\sqrt{{\mathcal{D}}_{p}(r)}}\left(-K(r) dt^{2}+dx_{\|}^{2}\right)+\sqrt{{\mathcal{D}}_{p}(r)}\left(\frac{dr^{2}}{K(r)}+r^{2}d\Omega_{8-p}^{2}\right) ,
\end{equation}
where ${\mathcal{D}}_{p}(r)$ is defined as
\begin{equation}\label{metric f2}
{\mathcal{D}}_{p}(r)=D_{p}(r)+\frac{2\sqrt{D_{p}(r)}}{\Omega_{8-p}r_{H}^{8-p}}\alpha\ln{\frac{\sqrt{D_{p}(r)}}{4G_{10-p}}}
+\frac{2\sqrt{D_{p}(r)}}{\Omega_{8-p}r_{H}^{8-p}}\eta e^{-\frac{\Omega_{8-p}\sqrt{D_{p}(r)}}{4G_{10-p}}}+\frac{32 G_{10-p}^{2}}{(\Omega_{8-p}r_{H}^{8-p})^{2}}\lambda,
\end{equation}
Here   $D_{p}(r)$ is defined by   (\ref{metric f}) and $K(r)$ is defined by (\ref{metric ff}). The area of the  event horizon   ($\mathcal{A}$) of the quantum corrected  geometry  can be described by the modified metric (\ref{metric2}). So, we can  write the  area for the quantum corrected geometry as
\begin{equation}\label{A2}
\mathcal{A}=\Omega_{8-p}r_{H}^{8-p}\sqrt{{\mathcal{D}}_{p}(r_{H})}.
\end{equation}
Now for the  constants $\alpha$, $\eta$, and $\lambda$,  we can expand above expression, and to  the first order approximation write this area as
\begin{equation}\label{A3}
\mathcal{A}=\Omega_{8-p}r_{H}^{8-p}\sqrt{D_{p}(r_{H})}Z,
\end{equation}
where
\begin{equation}
Z\approx1+\frac{\alpha\ln{\frac{\sqrt{D_{p}(r_{H})}}{4G_{10-p}}}}{\Omega_{8-p}r_{H}^{8-p}\sqrt{D_{p}(r_{H})}}
+\frac{\eta e^{-\frac{\Omega_{8-p}\sqrt{D_{p}(r_{H})}}{4G_{10-p}}}}{\Omega_{8-p}r_{H}^{8-p}\sqrt{D_{p}(r_{H})}}+\frac{16 G_{10-p}^{2}}{\Omega_{8-p}^{2}(r_{H}^{8-p})^{2}\sqrt{D_{p}(r_{H})}}\lambda
\end{equation}
Here the second order terms of the coefficients for the quantum  corrected geometry have been   neglected. Now, from using the fact that $D_{p}(r_{H})=\cosh{\beta}$ (from Eq.  (\ref{metric f})),  we can reproduce the entropy (\ref{S}) as,
\begin{equation}
S_{BH}=\frac{\mathcal{A}}{4G_{10-p}}.
\end{equation}
Thus, it is possible to use these quantum corrections to the thermodynamics of the black branes to construct the geometry of the quantum corrected black branes. It may be noted that the exact form of the non-perturbative  corrections cannot be directly obtained from modifying the action for this system. However, here we have argued using the  Jacobson formalism  \cite{gr12}, that  such non-perturbative  quantum corrections to the  geometry can be obtained from non-perturbative corrections to the thermodynamics of these black branes   \cite{gr14, gr16}.
\section{Conclusion}
In this paper, we have analyzed the thermodynamics of black branes at quantum scales. This will be done by analyzing the effects of both perturbative and non-perturbative corrections. The perturbative corrections correct this system by a logarithmic correction. Apart from this logarithmic correction term, we also analyze next the leading order perturbative  correction to the entropy of this system. The non-perturbative corrections occur in the form of the exponential function of the area.   These correction terms correct the specific heat of this system. This in turn corrects the thermodynamic stability of this system. Thus, we have analyzed the effects of perturbative and non-perturbative quantum corrections on the stability of $D_p$ branes.It is observed that the behavior of $D_6$ branes is different for other $D_p$ branes, with $p<6$.   We also calculate the corrections to the Helmholtz free energy and internal energy of this system. It is observed that the behavior of perturbative corrections is different from the behavior of non-perturbative corrections. We also used the  Jacobson formalism  \cite{gr12} to argue that these corrections to the thermodynamics should corrected the geometry of black branes. In fact, we explicitly obtained the corrections to the metric corresponding to both perturbative and non-perturbative  quantum corrections.

It will be interesting to investigate the effect of such perturbative and non-perturbative quantum corrections on different black holes. It may be noted that the effect of perturbative quantum corrections on various black holes has been analyzed \cite{40a, 40b, 40c, 40d}. However, the effect of non-perturbative quantum corrections on such black holes has not been discussed. As we have observed that such non-perturbative corrections can have important consequences for the stability of black branes, we expect that they will also have important consequences for the stability of different black holes.   This is because such non-perturbative quantum corrections will modify the relation between area and entropy, and this modification would in turn correct the specific heat of various black holes. This corrected specific heat would modify the stability of such black holes. Thus, we expect these non-perturbative corrections to have important consequences for the stability of various black holes. So, it would be interesting to investigate the effects of non-perturbative quantum corrections on AdS black holes. Such non-perturbative effects can also be investigated using the AdS/CFT correspondence.


\begin{thebibliography}{99}



\bibitem{1a}S. W. Hawking, Nature 248 (1974) 30.
\bibitem{2a}S. W. Hawking, Commun. Math. Phys. 43  (1975) 199.
\bibitem{4a}L. Susskind, J. Math. Phys. 36  (1995)  6377.
\bibitem{5a} R. Bousso, Rev. Mod. Phys. 74  (2002) 825.
\bibitem{6a}D. Bak and S. J. Rey, Class. Quant. Grav. 17  (2000) L1.
\bibitem{7a} S. K. Rama, Phys. Lett. B 457  (1999) 268.


\bibitem{18}
S. Hemming and L. Thorlacius,   JHEP 11 (2007) 086.
\bibitem{18a}R.~Gregory, S.~F.~Ross and R.~Zegers,
 JHEP  {09}  (2008) 029.
\bibitem{18b} J.~V.~Rocha, JHEP {08}  (2008) 075.
\bibitem{18c}Z.~H.~Li, B.~Hu and R.~G.~Cai, Phys. Rev. D {77} (2008)
104032.
\bibitem{18d} K.~Saraswat and N.~Afshordi,
 JHEP  136 (2020) {04}.

\bibitem{19}
R. B. Mann and S. N. Solodukhin,   Nucl. Phys. B 523 (1998) 293.
\bibitem{19a}A.~Sen, Gen. Rel. Grav. {44} (2012) 1947.


\bibitem{Ashtekar}
A. Ashtekar, \textit{Lectures on non-perturbative canonical gravity}, World Scientific: Singapore (1991).


\bibitem{Govindarajan}
T. R. Govindarajan, R. K. Kaul and V. Suneeta,  Class. Quantum Grav. 18 (2001) 2877.
\bibitem{29}
D. Birmingham and S. Sen, Phys. Rev. D 63 (2001) 047501.

\bibitem{32}
S. Das, P. Majumdar and R. K. Bhaduri,   Class. Quantum Grav. 19 (2002) 2355.
\bibitem{32a}S. Upadhyay, B. Pourhassan and H. Farahani,
  Phys. Rev. D 95 (2017) 106014.
\bibitem{32b} A.~Jawad, Class. Quant. Grav.  {37} (2020)   185020.
\bibitem{32c}B.~Pourhassan, Eur. Phys. J. C {79} (2019)   740.
\bibitem{32d} J.~Sadeghi, B.~Pourhassan and M.~Rostami,
Phys. Rev. D  {94} (2016) no.6, 064006.
\bibitem{gr12}T. Jacobson, Phys. Rev. Lett. 75 (1995) 1260.
\bibitem{gr14} M.~Faizal, A.~Ashour, M.~Alcheikh, L.~Alasfar, S.~Alsaleh and A.~Mahroussah, Eur. Phys. J. C {77} (2017)  608.
\bibitem{gr16} F.~Hammad and M.~Faizal, Int. J. Mod. Phys. D  {25}   (2016) 1650080

\bibitem{40a}B.~Pourhassan, S.~Dey, S.~Chougule and M.~Faizal,
 Class. Quant. Grav.  {37} (2020) 135004.
\bibitem{40b}B.~Pourhassan, S.~Upadhyay, H.~Saadat and H.~Farahani,
 Nucl. Phys. B {928} (2018) 415.
\bibitem{40c}B.~Pourhassan, M.~Faizal, Z.~Zaz and A.~Bhat,
 Phys. Lett. B  {773} (2017) 325.
\bibitem{40d} B.~Pourhassan, A.~Ovgun and I.~Sakalli,
 Int. J. Geom. Meth. Mod. Phys.  {17} (2020)  2050156
\bibitem{point4}S-W. Wei  and Y-X. Liu
Phys. Rev. Lett. 115  (2015) 111302
\bibitem{point5} S-W. Wei, Y-X. Liu and R. B. Mann
Phys. Rev. Lett. 123 (2019) 071103



\bibitem{2007.15401}
A. Chatterjee and A. Ghosh,   Phys. Rev. Lett. 125 (2020) 041302.
\bibitem{Dabholkar}
A. Dabholkar, J. Gomes and S. Murthy,   JHEP 03 (2015) 074.

\bibitem{ds12}A.~Dabholkar, J.~Gomes and S.~Murthy, JHEP 04  (2013) 062.
\bibitem{ds14} S. Murthy and B. Pioline,  JHEP 09 (2009) 022.
\bibitem{ca15} Z. Xiao and D. Zhou, JHEP  028 (2015) 1509.
\bibitem{ca16} D. Zhou and Z. Xiao,   JHEP  134 (2015)   1507.
\bibitem{da12}J. X. Lu and R. Wei,   JHEP  100 (2013) 1304.
\bibitem{da14}J. X. Lu, R. Wei and J. Xu,  JHEP  012 (2012) 1212.

\bibitem{9412184}
M. J. Duff, R. R Khuri and J. X. Lu, Phys. Rept. 259  (1995) 213.

\bibitem{Peet:2000hn}
A.~W.~Peet,  hep-th/0008241 [hep-th]

\bibitem{x1}
S.~Mukherji and S.~S.~Pal,  JHEP 0205, 026 (2002)
\bibitem{x2}
J.~E.~Lidsey, S.~Nojiri, S.~D.~Odintsov and S.~Ogushi,   Phys. Lett. B 544 (2002) 337.
\bibitem{x21}
M.~H.~Dehghani and A.~Khoddam-Mohammadi, Phys. Rev. D 67 (2003) 084006.
\bibitem{x4}
S.~Das and V.~Husain,   Class. Quantum Grav.  20 (2003) 4387.


\end{thebibliography}
\end{document}